\documentclass[12pt,english,american]{article} 
\usepackage[letterpaper]{geometry}
\usepackage[normalem]{ulem}
\usepackage{color,bm}
\usepackage{babel}
\usepackage{amsmath,amssymb,graphicx}
\usepackage{setspace}
\usepackage[authoryear]{natbib}
\usepackage{cancel}
 
\usepackage[unicode=true,pdfusetitle,
bookmarks=true,bookmarksnumbered=false,bookmarksopen=false,
	breaklinks=false,pdfborder={0 0 1},backref=false,colorlinks=false]
	{hyperref}
\makeatletter\@ifundefined{date}{}{\date{}}
 
\AtBeginDocument{}
\newtheorem{thm}{Theorem}
\newtheorem{rem}{Remark}
\newtheorem{exm}{Example}
\makeatother

\begin{document}

\title{Multi-Resolution Spatial Random-Effects Models for Irregularly Spaced Data}

\date{April 21, 2015}

\author{ShengLi Tzeng\\
 Department of Public Health\\
 China Medical University\\
 Taichung 40402, Taiwan\\
 \emph{slt.cmu@gmail.com} \and Hsin-Cheng Huang\\
 Institute of Statistical Science\\
 Academia Sinica\\
 Taipei 11529, Taiwan\\
 \emph{hchuang@stat.sinica.edu.tw} }

\maketitle

\begin{abstract}
The spatial random-effects model is flexible in modeling spatial covariance functions,
and is computationally efficient for spatial prediction via fixed rank kriging.
However, the success of this model depends on an appropriate set of basis functions.
In this research, we propose a class of basis functions extracted from thin-plate splines.
These functions are ordered in terms of their degrees of smoothness with a higher-order function corresponding to larger-scale features
and a lower-order one corresponding to smaller-scale details,
leading to a parsimonious representation for a nonstationary spatial covariance function. Consequently,
only a small to moderate number of functions are needed in a spatial random-effects model.
The proposed class of basis functions has several advantages over commonly used ones.
First, we do not need to concern about the allocation of the basis functions,
but simply select the total number of functions corresponding to a resolution.
Second, only a small number of basis functions is usually required, which facilitates computation.
Third, estimation variability of model parameters can be considerably reduced,
and hence more precise covariance function estimates can be obtained.
Fourth, the proposed basis functions depend only on the data locations but not the measurements taken at those locations,
and are applicable regardless of whether the data locations are sparse or irregularly spaced.
In addition, we derive a simple close-form expression for the maximum likelihood estimates of
model parameters in the spatial random-effects model.
Some numerical examples are provided to demonstrate the effectiveness of the proposed method.
\medskip
	
\end{abstract}

\section{Introduction}

Consider a sequence of independent spatial processes, $\{y(\bm{s},t):\bm{s}\in D\}$;
$t=1,\dots,T$, defined on a $d$-dimensional spatial domain $D\subset\mathbb{R}^{d}$.
The processes are assumed to have mean $\mu(\bm{s},t)$ and a common spatial
covariance function $C(\bm{s},\bm{s}^{*})=\mathrm{cov}(y(\bm{s},t),\, y(\bm{s}^{*},t))$, for $t=1,\dots,T$.
Suppose that we observe data $\bm{z}_{t}\equiv\left(z(\bm{s}_{1},t),\dots,z(\bm{s}_{n},t)\right)'$;
$t=1,\dots,T$, at $n$ distinct locations, $\bm{s}_{1},\dots,\bm{s}_{n}\in D$, with
additive white noise $\bm{\varepsilon}_{t}$ according to 
\begin{equation}
\bm{z}_{t}=\bm{y}_{t}+\bm{\varepsilon}_{t};\quad t=1,\dots,T,
\label{eq:uni-obs-equation-z}
\end{equation}

\noindent where $\bm{y}_t=(y(\bm{s}_1,t),\dots,y(\bm{s}_n,t))'$, $\bm{\varepsilon}_t\sim N(\bm{0},\sigma_\epsilon^2\bm{I}_n)$
is uncorrelated with $\bm{y}_t$, and $\bm{\varepsilon}_t$'s are mutually uncorrelated.
The goal is to estimate $C(\cdot,\cdot)$ and predict $y(\cdot,t)$; $t=1,\dots,T$, based on $\bm{z}_1,\dots,\bm{z}_T$
without imposing a stationary assumption or a parametric structure.

We consider the spatial random-effects model (e.g., \citealp{cressie2008fixed[clm]}; \citealp{wikle2010[clm]};
\citealp{lemos2012conditionally[clm-review]}):
\begin{align}
	y(\bm{s},t)
=&~ \mu(\bm{s},t)+ \bm{w}_{t}'\bm{f}(\bm{s})+\xi(\bm{s},t)\notag\\
=&~ \mu(\bm{s},t) +\sum_{k=1}^{K}w_{k}(t)f_{k}(\bm{s})+\xi(\bm{s},t);\quad \bm{s}\in D,\, t=1,\dots,T,
\label{eq:uni-general-form-y}
\end{align}

\noindent where $f_k(\cdot)$'s are pre-specified basis functions with $K\leq n$,
$\bm{f}(\bm{s})=(f_{1}(\bm{s}),\dots,f_{K}(\bm{s}))'$,
$\bm{w}_{t}=(w_{1}(t),\dots,w_{K}(t))'\sim N(\bm{0},\bm{M})$;
$t=1,\dots,T$, are random effects, and $\xi(\bm{s},t)\sim N(0,\sigma_{\xi}^2)$ is a white-noise
process. Here $\bm{w}_{t}$'s and $\xi(\bm{s},t)$'s are mutually uncorrelated. This model is
flexible for modeling stationary or nonstationary spatial covariance functions and can produce fast prediction (e.g., \citealp{wikle2010[clm]}).
The spatial covariance function is
\begin{equation}
C(\bm{s},\bm{s}^*)=\mathrm{cov}(y(\bm{s},t),y(\bm{s}^*,t))=\bm{f}(\bm{s})'\bm{M}\bm{f}(\bm{s}^*)
+\sigma_{\xi}^2 I(\bm{s}=\bm{s}^*);\quad\bm{s},\bm{s}^*\in D.
\label{eq:covariance}
\end{equation}
\noindent

Given $\{f_1(\cdot),\dots,f_K(\cdot)\}$, the model \eqref{eq:uni-general-form-y} depends only on the
parameters $\bm{M}$ and $\sigma_{\xi}^{2}$. Many approaches have been proposed to estimate these parameters, including
a  method of moments (\citealp{cressie2008fixed[clm]})
and maximum likelihood (\citealp{katzfuss2009EM[clm]}).
Commonly used basis functions include radial basis functions (e.g., \citealp{cressie2008fixed[clm]} and \citealp{nychka2014multi}),
discrete kernel basis functions (e.g.,  \citealp{barry1996blackbox} and \citealp{wikle2010[clm]}), and wavelets
(e.g., \citealp{nychka2002} and \citealp{shi2007global}). Although wavelet basis functions are advantageous to have multi-resolution features,
they are mainly restricted for data observed on a regular grid with no (or few) missing observations.
In general, different basis functions work well under different situations. However,
how to select and allocate the basis functions (e.g., centers and radii) is an art and has rarely been discussed in the literature. 

In what follows, we provide some examples showing how estimation of $\bm{M}$ and $\sigma_{\xi}^2$, and thus $C(\cdot.\cdot)$,
is affected by the choice of the following bisquare (radial) basis functions:
\begin{equation}
f_{k}(\bm{s})=\bigg(1-\frac{\|\bm{s}-\bm{b}_{k}\|^2}{r_k^2}\bigg)^2 I(\|\bm{s}-\bm{b}_{k}\|<r_{k}),
\label{eq:1D example}
\end{equation}

\noindent which is centered at $\bm{b}_{k}$ and has a local bounded support
$\{\bm{s}\in\mathbb{R}^d:\|\bm{s}-\bm{b}_k\|<r_k\}$ controlled by a radius $r_k$, for $k=1,\dots,K$.

\begin{figure}\centering
\includegraphics[scale=0.14]{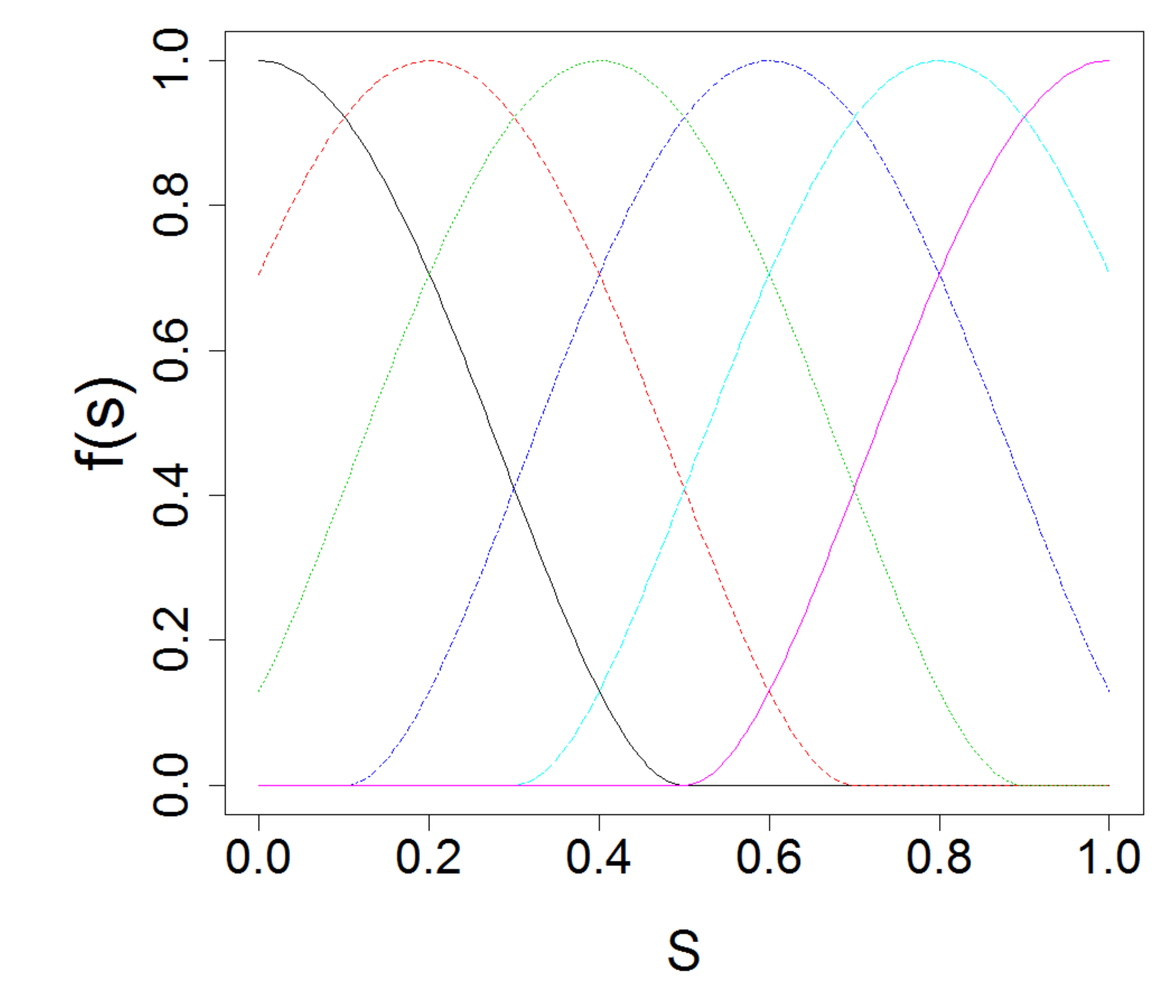}  \hspace{2cm} \includegraphics[scale=0.14]{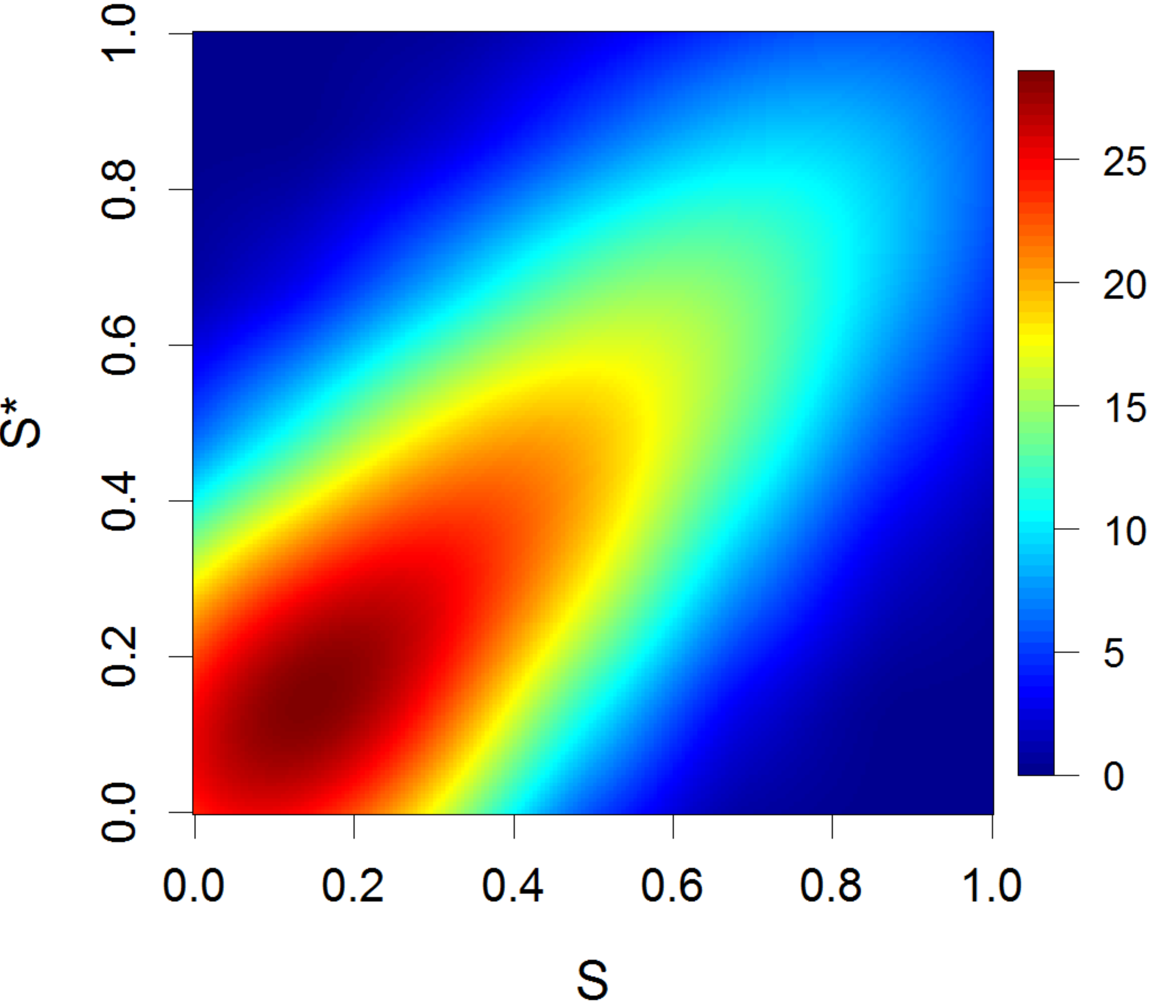}\vspace{-0.1cm}\\
(a1) \hspace{6.05cm} (a2)\smallskip\\
\includegraphics[scale=0.14]{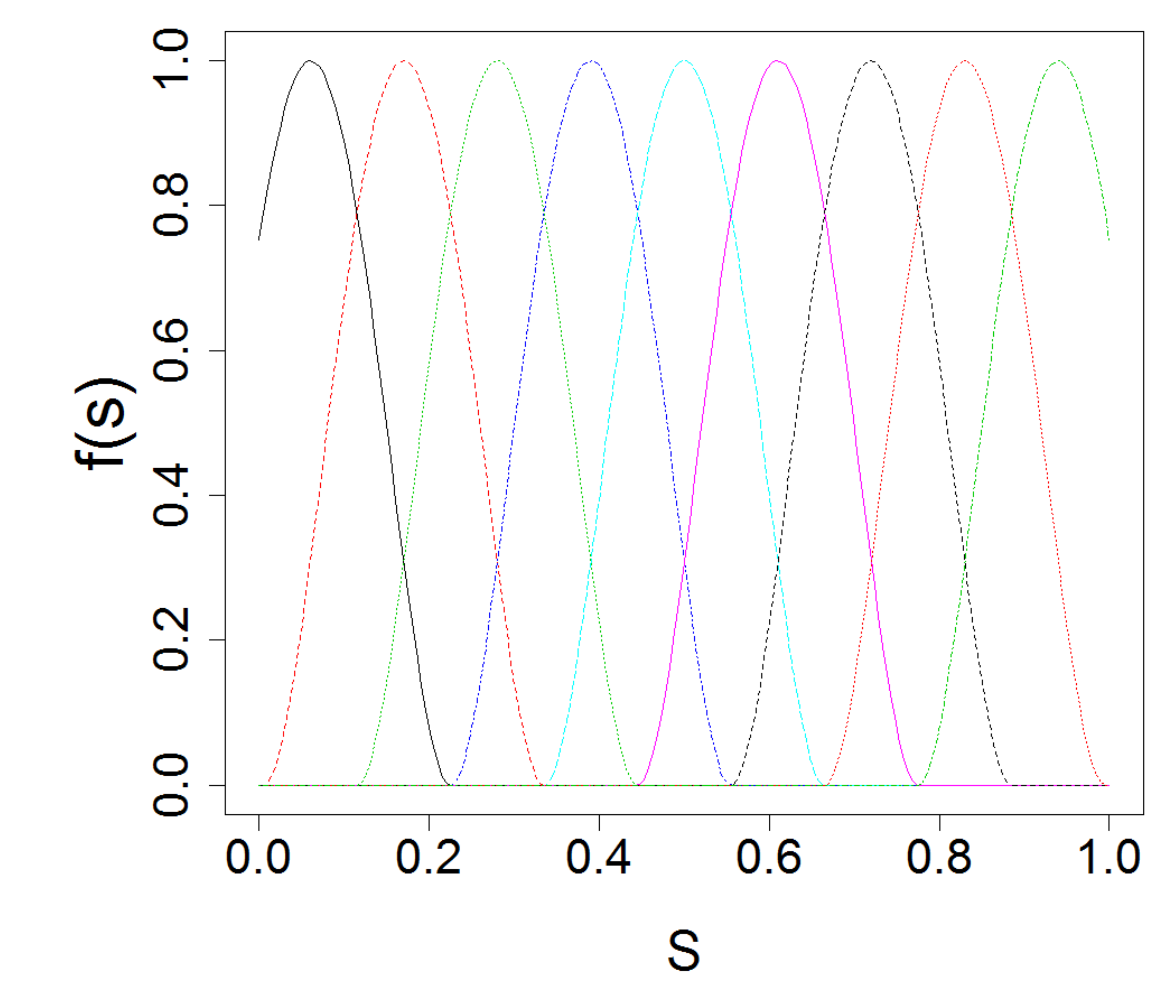}  \hspace{2cm} \includegraphics[scale=0.14]{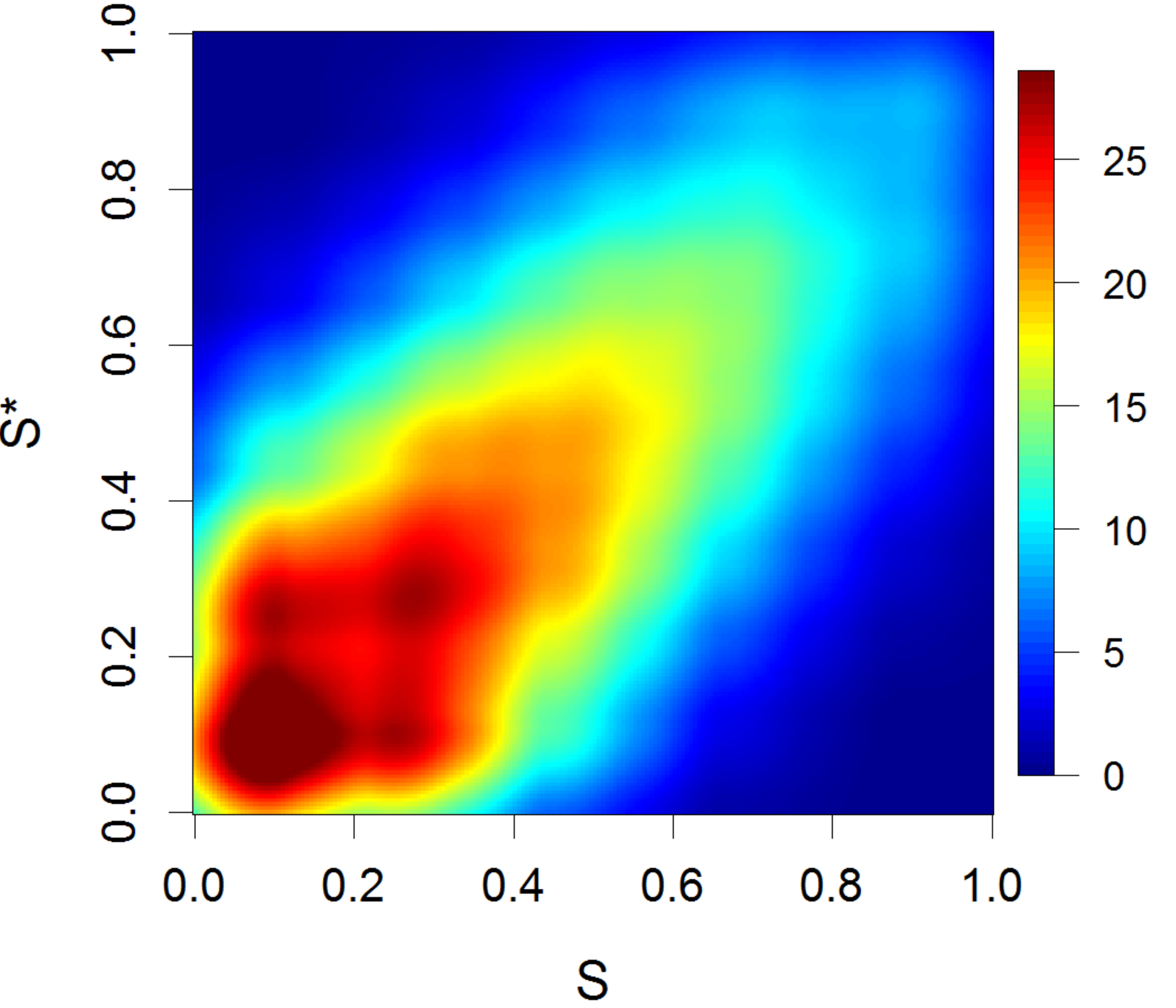}\vspace{-0.1cm}\\
(b1) \hspace{6.05cm} (b2)\smallskip\\
\includegraphics[scale=0.14]{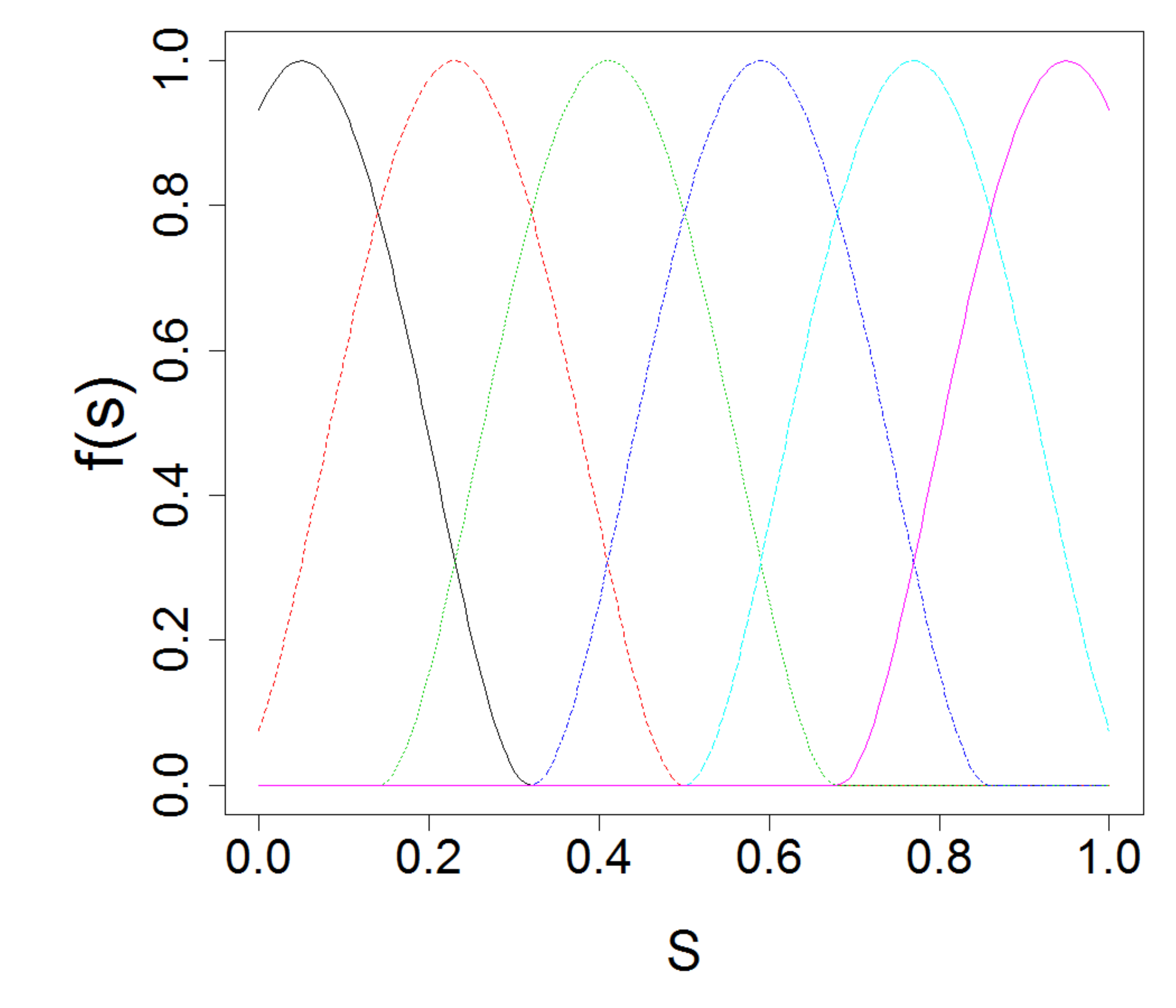} \hspace{2cm} \includegraphics[scale=0.14]{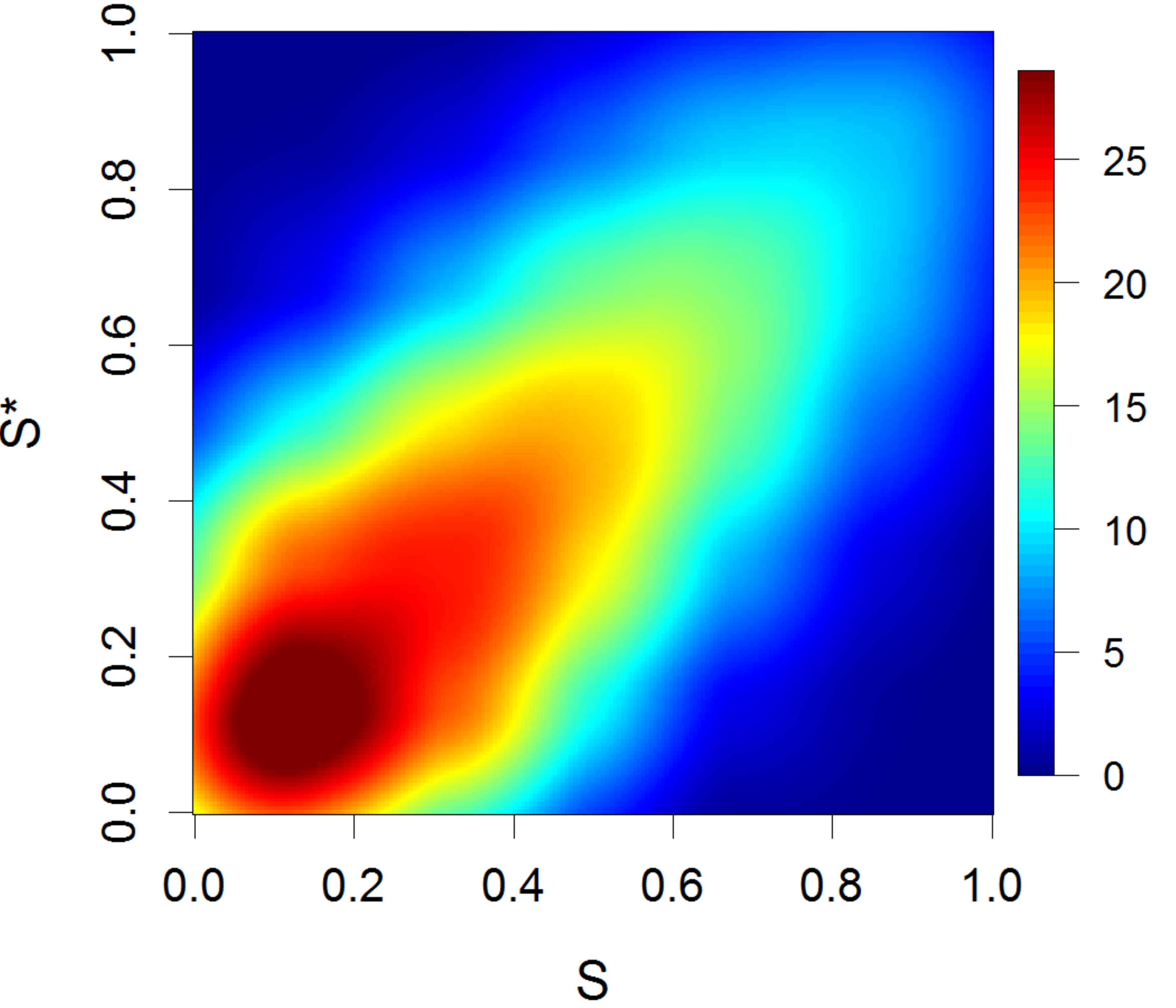}\vspace{-0.1cm}\\
(c1) \hspace{6.05cm} (c2)\smallskip\\
\includegraphics[scale=0.14]{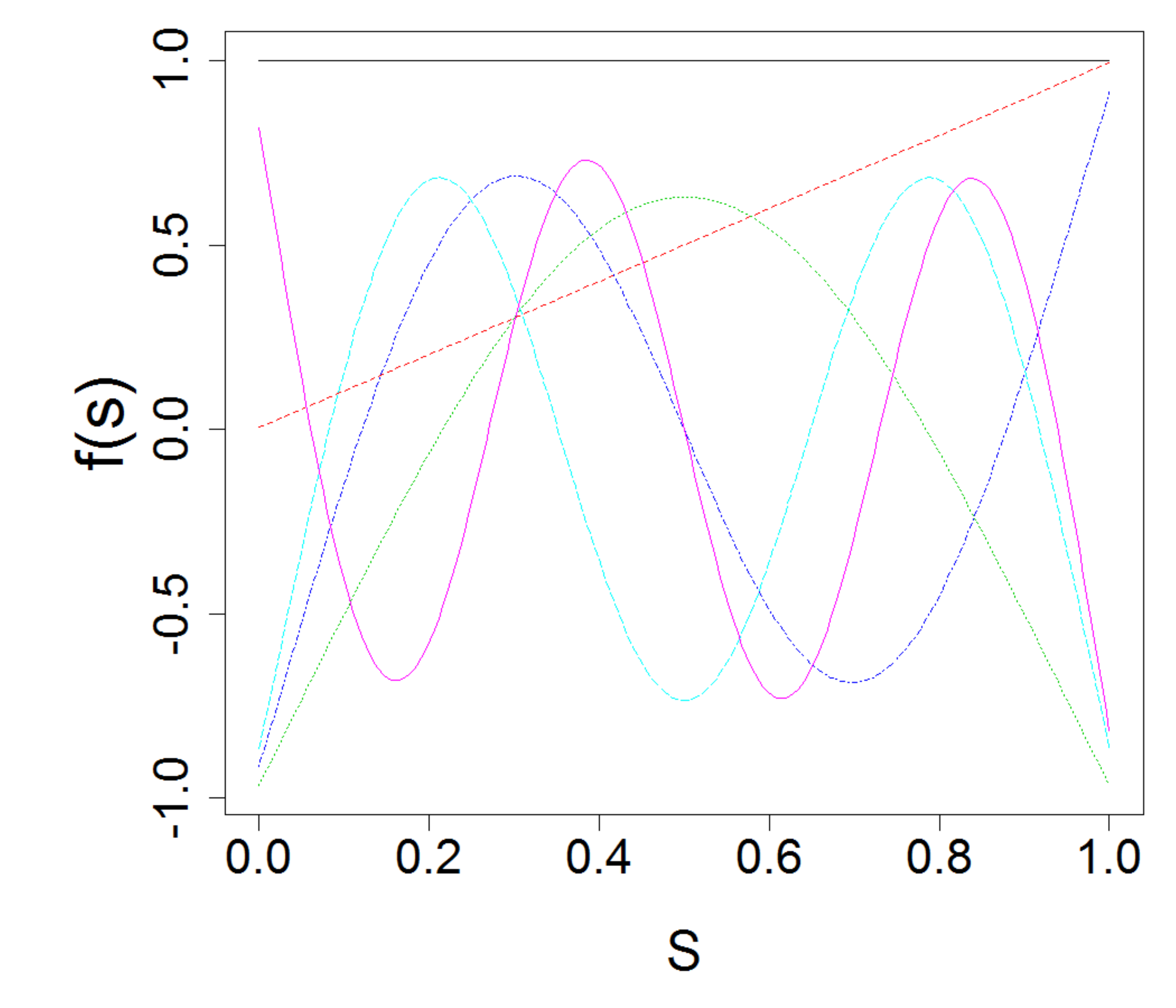} \hspace{2cm} \includegraphics[scale=0.14]{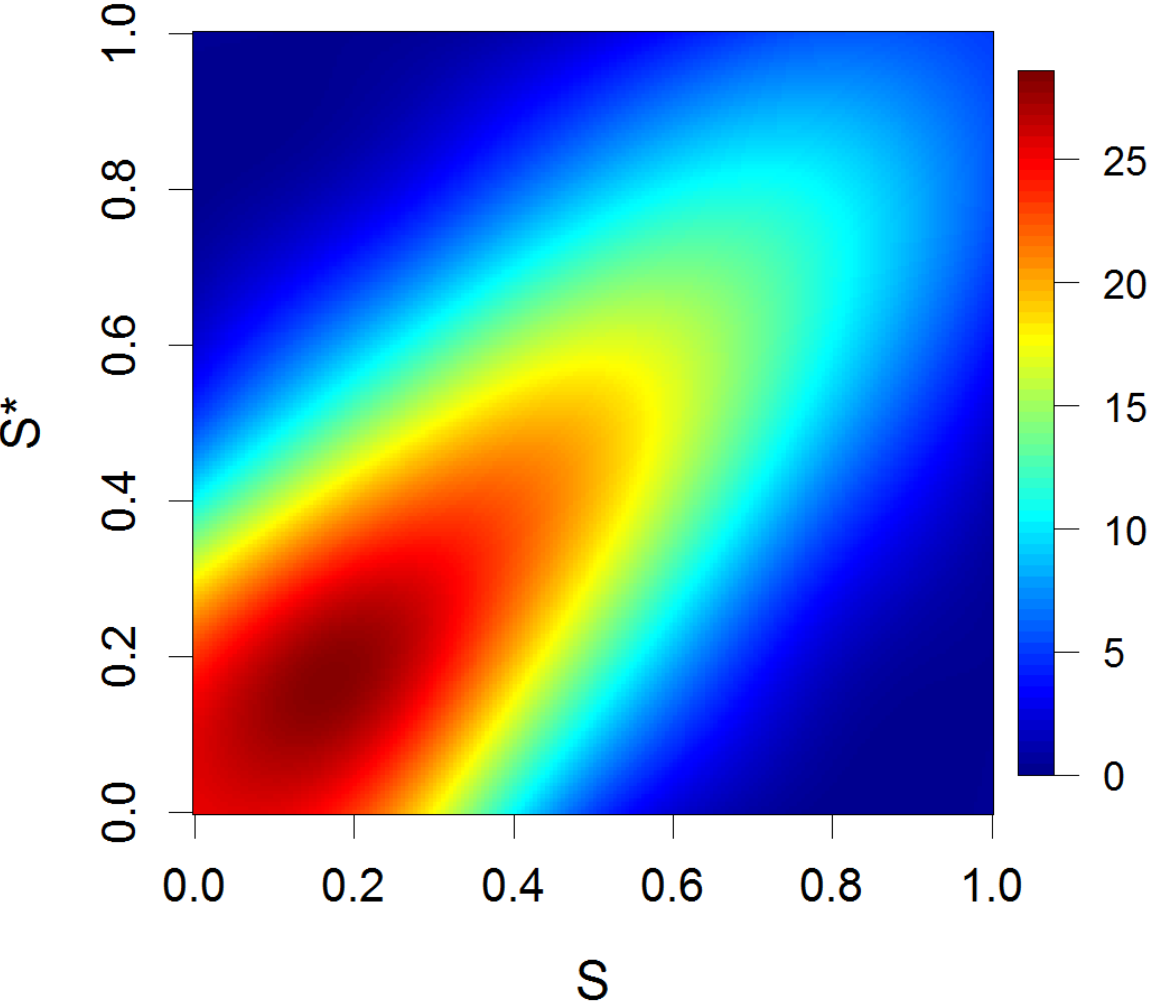}\vspace{-0.1cm}\\
(d1) \hspace{6.05cm} (d2)
\caption{(a1) Six basis functions corresponding to $\bm{f}^{(0)}(\cdot)$; (a2) The true spatial covariance function;
(b1) Nine basis functions corresponding to $\bm{f}^{(1)}(\cdot)$; (b2) Spatial covariance function obtained from $\bm{f}^{(1)}(\cdot)$;
(c1) Six basis functions corresponding to $\bm{f}^{(2)}(\cdot)$; (c2) Spatial covariance function obtained from $\bm{f}^{(2)}(\cdot)$;
(d1) Six basis functions from the proposed method; (d2) Spatial covariance function obtained from the six proposed basis functions. }
\label{fig:motivation-part1} 
\end{figure}

\begin{exm}
Assume that the underlying covariance function is given by the
spatial random-effects model of \eqref{eq:uni-general-form-y} with $D=[0,1]$, $K=6$, $\bm{M}=\mathrm{diag}(17,14,11,8,5,2)$,
$\sigma_{\xi}^2=0$, and $f_k^{(0)}(\cdot)$'s given by \eqref{eq:1D example} (see Figure~\ref{fig:motivation-part1} (a1)), where
$\bm{b}_{k}=0.2(k-1)$; $k=1,\dots,6$ and $r_1=\cdots=k_6=0.5$.
Then the spatial covariance function is $C^{(0)}(\bm{s},\bm{s}')=\bm{f}^{(0)}(\bm{s})'\bm{M}\bm{f}^{(0)}(\bm{s}^*)$
(Figure~\ref{fig:motivation-part1} (a2)), where $\bm{f}^{(0)}(\bm{s})=(f_{1}^{(0)}(\bm{s}),\dots,f_6^{(0)}(\bm{s}))'$.
\label{ex1}
\end{exm}

To mimic a situation in practice, instead of approximating $C^{(0)}(\cdot,\cdot)$ in Example \ref{ex1}
using $\bm{f}^{(0)}(\cdot)$, we consider a different set of bisque basis functions,
$\bm{f}^{(1)}(\bm{s})=(f_{1}^{(1)}(\bm{s}),\dots,f_{9}^{(1)}(\bm{s}))'$, formed by
$\bm{b}_{k}=0.11(k-1)+0.06$; $k=1,\dots,9$ and $r_1=\cdots=r_9=0.165$ (Figure~\ref{fig:motivation-part1} (b1)).
Let $\bm{M}^{(1)}$ be the optimal matrix that minimizes the integrated squared error $\mathrm{ISE}\big(\bm{f}^{(1)},\bm{M}\big)$
over all non-negative definite $9\times 9$ matrix $\bm{M}$, where 
\begin{equation}
\mathrm{ISE}\big(\bm{f},\bm{M}\big)=
\int_{D}\int_{D}\big\{\bm{f}(\bm{s})'\bm{M}\bm{f}(\bm{s}^*)-C^{(0)}(\bm{s},\bm{s}^*)\big\}^{2}d\bm{s}\,d\bm{s}^{*}.
\label{eq:ISE 1D}
\end{equation}

\noindent Then the covariance function that has the smallest ISE based on $\bm{f}^{(1)}(\cdot)$
is $C^{(1)}(\bm{s},\bm{s}^*)=\bm{f}^{(1)}(\bm{s})'\bm{M}^{(1)}\bm{f}^{(1)}(\bm{s}^*)$ (Figure \ref{fig:motivation-part1}~(b2)).
The approximation can be seen to be poor, because $\bm{b}_k$'s and $r_k$'s are not well chosen,
despite that a larger number of basis functions are used and the approximation involves no estimation error.

Now consider another set of bisquare basis functions,
$\bm{f}^{(2)}(\bm{s})=(f_{1}^{(2)}(\bm{s}),\dots,f_{6}^{(2)}(\bm{s}))'$ to approximation $C^{(0)}(\cdot,\cdot)$,
where $\bm{b}_{k}=0.18(k-1)+0.05$; $k=1,\dots,6$ and $r_1=\cdots=r_6=0.27$ (see Figure \ref{fig:motivation-part1}~(c1)).
Here $r_{k}$'s are determined by $1.5$ times the minimal distance between $\bm{b}_{k}$'s as suggested by \citet{cressie2008fixed[clm]}.
Similar to $C^{(1)}(\cdot,\cdot)$, the best covariance function based on $\bm{f}^{(2)}(\cdot)$
is $C^{(2)}(\bm{s},\bm{s}^*)=\bm{f}^{(2)}(\bm{s})'\bm{M}^{(2)}\bm{f}^{(2)}(\bm{s}^*)$ (Figure \ref{fig:motivation-part1}~(c2)).
Although $C^{(2)}(\cdot,\cdot)$ is smoother than $C^{(1)}(\cdot,\cdot)$, it produces a larger bias. 
Clearly, the quality of approximation highly depends on the choice of $K$, $\bm{b}_{k}$'s and $r_k$'s.

Instead of selecting $\bm{b}_k$'s and $r_k$'s for the bisquare functions of \eqref{eq:1D example},
we shall propose a new class of basis functions, which involves no selection of centers and radii,
and are ordered in terms of their degrees of smoothness.
Figure \ref{fig:motivation-part1} (d1) shows a class of $K=6$ basis functions obtained from our method, which will be introduced
in Section~\ref{sec:basis definition}. The covariance function based on this class of functions is shown in
Figure \ref{fig:motivation-part1} (d2). Comparing it to $C^{(1)}(\cdot,\cdot)$ and $C^{(2)}(\cdot,\cdot)$,
a significant improvement can be seen even though only 6 functions are used.

\begin{figure}
\begin{tabular}{cc}
\includegraphics[scale=0.28]{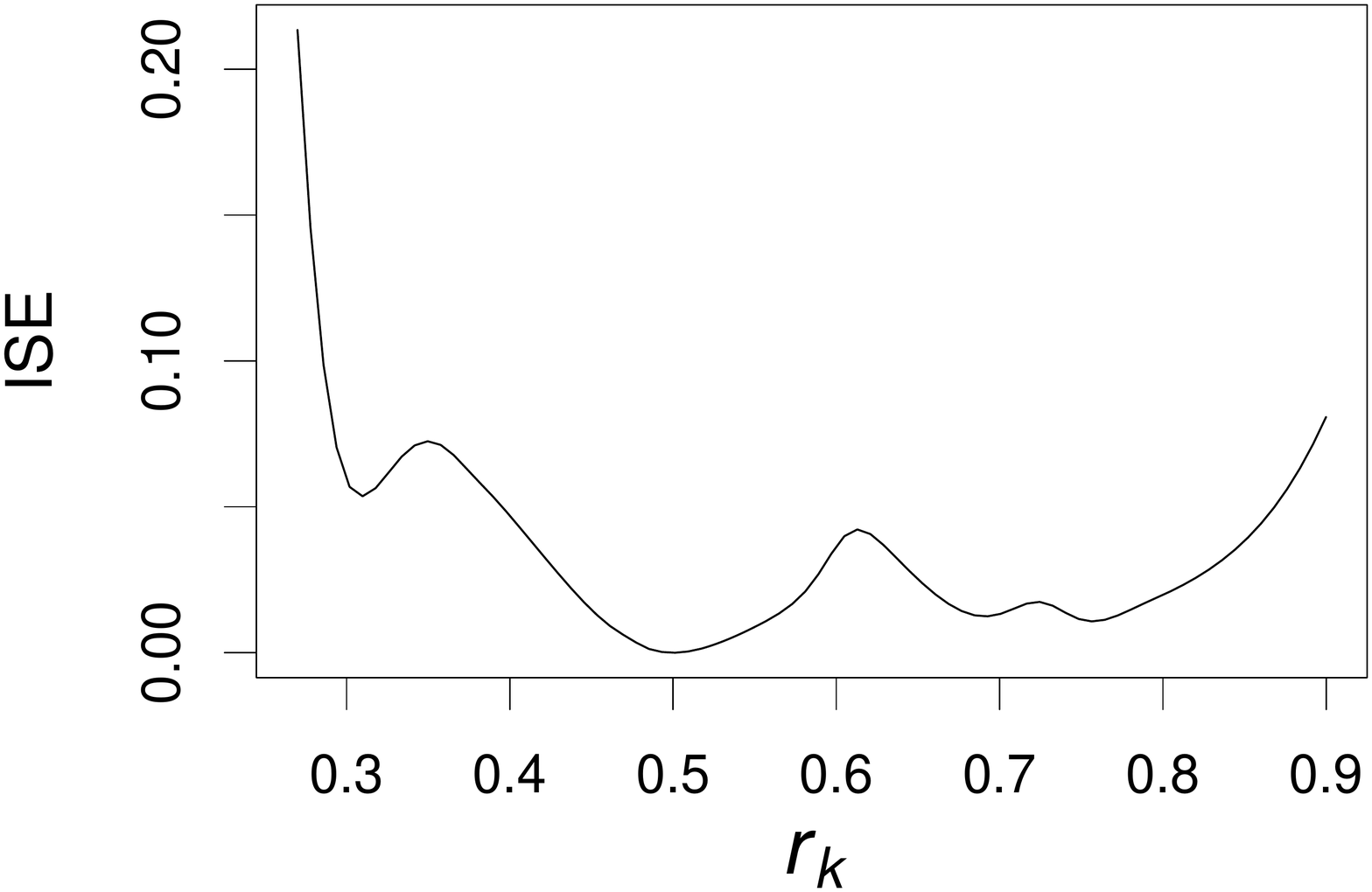} & \includegraphics[scale=0.28]{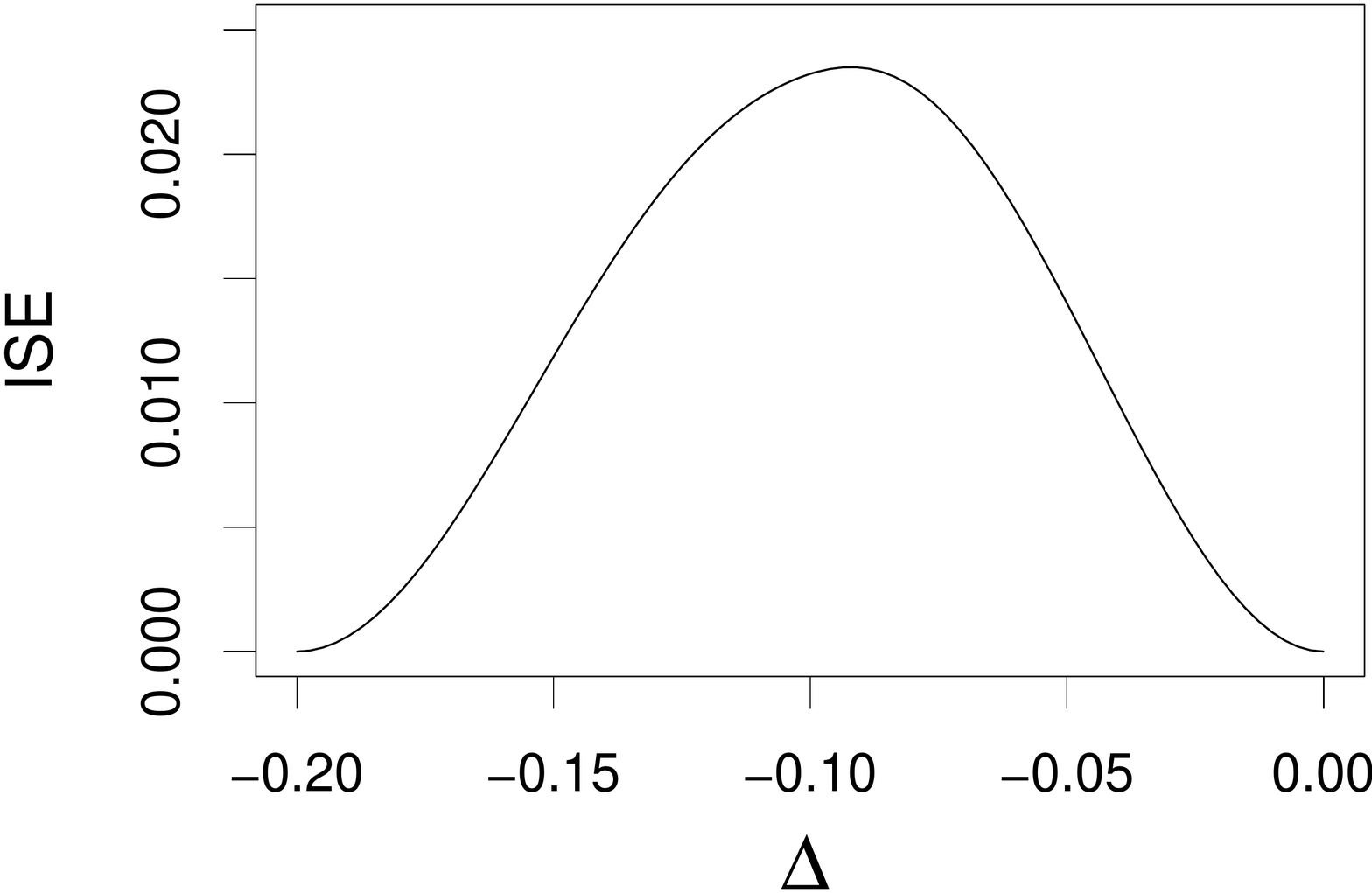}\vspace{-0.5cm}\\
(a) & (b)\tabularnewline
\end{tabular}
\caption{(a) ISE values with respect to  $r_k$ based on six basis functions of \eqref{eq:1D example};
(b) ISE values with respect to $\Delta$ with $\bm{b}_k=0.2(k-1)+\Delta$ based on seven basis functions of \eqref{eq:1D example}.}
\label{fig:motivation-part2} 
\end{figure}

To further investigate the effect of $\bm{b}_{k}$'s and $r_k$'s in covariance function estimation, we consider two additional examples.
For the first example, we apply the same basis functions of $\bm{f}^{(0)}(\bm{s})$ except that $r_1=\cdots=r_6=r\in[0.25,0.9]$.
Figure \ref{fig:motivation-part2}~(a) shows how the ISE of \eqref{eq:ISE 1D} varies as a function of $r$. Not surprisingly,
covariance function estimation is highly affected by $r$.
For the second example, we consider the same bisque functions of \eqref{eq:1D example}
with $\bm{b}_{k}=0.2(k-1)+\Delta$; $k=1,\dots,7$ and $r_1=\cdots=r_7=0.5$, similar to those in Example \ref{ex1}.
These can be regarded as shifted versions of $\bm{f}^{(0)}(\bm{s})$ controlled by a shift parameter $\Delta$.
Figure \ref{fig:motivation-part2}~(b) shows the ISE of \eqref{eq:ISE 1D} with respect to $\Delta\in[-0.2,0]$.
While ISE is less affected by $\Delta$ than $r$ in the first example, a poorly chosen $\Delta$ can still cause some
significant bias in covariance function estimation.

In this research, we propose a class of basis functions extracted from thin-plate splines. These functions are ordered in terms of their degrees of smoothness with a higher-order function corresponding to larger-scale features
and a lower-order one corresponding to smaller-scale details,
leading to a parsimonious representation for a nonstationary spatial covariance function. Consequently,
only a small to moderate number of functions are needed in a spatial random-effects model.
The proposed class of basis functions has several advantages over commonly used ones.
First, we do not need to concern about the allocation of the basis functions,
but simply select the total number of functions corresponding to a resolution.
Second, only a small number of basis functions is usually required, which facilitates computation.
Third, estimation variability of model parameters can be considerably reduced,
and hence more precise covariance function estimates can be obtained.
Fourth, the proposed basis functions depend only on the data locations but not the measurements taken at those locations,
and are applicable regardless of whether the data locations are sparse or irregularly spaced.

The rest of the article is organized as follows. Section 2 introduces the proposed class of basis functions.
In Section 3, we apply the proposed basis functions to spatial random-effects models,
and derive simple close-form expressions for the maximum likelihood estimates of the model parameters.
Some simulation examples and an application to a daily-temperature dataset in Canada are presented in Section 4.

\section{The Proposed Ordered Set of Basis Functions\label{sec:basis definition}}

The proposed class of basis functions will be developed using thin-plate splines (TPSs).
We shall first provide some basic knowledge about TPS.
Given noisy data $\bm{Z}_1,\dots,\bm{Z}_n$ observed at $n$ distinct control points, $\bm{s}_1,\dots,\bm{s}_n\in\mathbb{R}^d$,
a TPS function $f(\bm{s})$; $\bm{s}\in\mathbb{R}^d$, can be obtained by minimizing
\begin{equation}
\sum_{i=1}^{n}(Z_i-f(\bm{s}_i))^2+\rho J(f),
\label{eq:TPS smoothing}
\end{equation}

\noindent where $\bm{s}=(x_1,\dots,x_d)'$,
\begin{equation}
J(f)=\int_{\mathbb{R}^{d}}\sum_{\nu_{1}+\cdots+\nu_{d}=2}\frac{2!}{\nu_{1}!\cdots\nu_{d}!}
\left(\frac{\partial^{2}f(\bm{s})}{\partial x_1^{\nu_1}\cdots\partial x_d^{\nu_d}}\right)^2 d\bm{s}\:\geq 0,
\label{eq:roughness}
\end{equation}

\noindent is a smoothness penalty, and $\rho\geq 0$ is a tuning parameter. 
It is known that (e.g., \citealp{wahba1980some[TPS]};  \citealp{green1993nonparametric}) for $\rho>0$,
the solution of \eqref{eq:TPS smoothing} satisfies
\begin{equation}
f(\bm{s})=\bm{\alpha}'\bm{\phi}(\bm{s})+\beta_{0}+\sum_{j=1}^{d}\beta_{j}x_{j}\textrm{ subject to }\bm{X}'\bm{\alpha}=\bm{0},
\label{eq:TPS definition}
\end{equation} 
where $\bm{s}_i=(x_{i1},\dots,x_{id})'$; $i=1,\dots,n$,
\begin{equation}
\bm{X}=\left(
\begin{matrix}
1 & x_{11} & \cdots & x_{1d}\\
\vdots && \ddots & \\
1 & x_{n1} & \cdots & x_{nd} 
\end{matrix}
\right),
\label{eq:linear polynomials}	
\end{equation}
and $\bm{\phi}(\bm{s})=(\phi_1(\bm{s}),\dots,\phi_n(\bm{s}))'$ with
\begin{equation}
\phi_i(\bm{s})=
\begin{cases}
\displaystyle\frac{1}{12}\|\bm{s}-\bm{s}_{i}\|^{3}; & \mathrm{if\:} d=1,\\
\displaystyle\frac{1}{8\pi}\|\bm{s}-\bm{s}_{i}\|^{2}\log\left(\|\bm{s}-\bm{s}_{j}\|\right); & \mathrm{if\:} d=2,\\
\displaystyle\frac{-1}{8}\|\bm{s}-\bm{s}_{i}\|; & \mathrm{if\:} d=3.
\end{cases}
\label{eq:elements in Phi}
\end{equation} 

\noindent A function $f(\bm{s})$ in the form of (\ref{eq:TPS definition}) is called a natural TPS function.
It has been shown that (e.g., Theorem 7.1 in \citealp{green1993nonparametric}) 
\begin{equation}
J(f)=\bm{\alpha}'\bm{\Phi}\bm{\alpha},
\label{eq:quadratic loss}
\end{equation}
where $\bm{\Phi}$ is the $n\times n$ matrix with the $(i,j)$-th element $\phi_{j}(\bm{s}_{i})$.

Assume that $\mathrm{rank}(\bm{X})=d+1$. We shall introduce our basis functions from the
natural TPS function space:
\begin{equation}
\mathcal{F}=\Big\{f(\cdot):f(\bm{s})=\bm{\alpha}'\bm{\phi}(\bm{s})+\beta_0+\sum_{j=1}^d\beta_j x_j,\,
\bm{\alpha}\in\mathbb{R}^n,\bm{\beta}\in\mathbb{R}^{d+1},\,\bm{X}'\bm{\alpha}=\bm{0}\Big\},
\label{eq:TPS space}
\end{equation}

\noindent where $\bm{\beta}=(\beta_0,\beta_1,\dots,\beta_d)'$. The proposed basis functions form a basis of $\mathcal{F}$,
and are defined as
\begin{equation}
f_{k}(\bm{s})=
\begin{cases}
1; & \quad k=1,\\
x_{k-1}; & \quad k=2,\dots,d+1,\\
\lambda_{k-d-1}^{-1}\big\{\bm{\phi}(\bm{s})-\bm{\Phi}'\bm{X}(\bm{X}'\bm{X})^{-1}\bm{x}\big\}'
\bm{v}_{k-d-1}\big\}; & \quad k=d+2,\dots,n,
\end{cases}\label{eq:basis detail}
\end{equation}

\noindent where $\bm{x}=(1,\bm{s}^{\prime})^{\prime} =(1,x_{1},\dots,x_{d})'$, $\bm{v}_k$ is the $k$-th column of $\bm{V}$,
$\bm{V}\mathrm{diag}(\lambda_1,\dots,\lambda_n)\bm{V}'$ is the eigen-decomposition of $\bm{Q\Phi Q}$ with $\lambda_1\geq\cdots
\geq\lambda_n$, and $\bm{Q}=\bm{I}-\bm{X}(\bm{X}'\bm{X})^{-1}\bm{X}'$.
Note that $\bm{\alpha}'\bm{\Phi}\bm{\alpha}>0$ for all $\bm{\alpha}\neq\bm{0}$ with $\bm{X}'\bm{\alpha}=\bm{0}$
(see Section 4 of \cite{Micchelli1986}).
Consequently, $\bm{a}'\bm{Q}\bm{\Phi}\bm{Q}\bm{a}>0$ for all $\bm{a}$ satisfying $\bm{Q}\bm{a}\neq\bm{0}$, which implies
$\mathrm{rank}(\bm{Q\Phi Q})=\mathrm{rank}(\bm{Q})=n-d-1$. Thus $\lambda_1\geq\cdots\geq\lambda_{n-d-1}>0$,
and hence $f_{d+2}(\cdot),\dots,f_n(\cdot)$ are well defined.

The following theorem gives some important properties of these basis functions with its proof given in Appendix.

\begin{thm}
Consider $f_k(\cdot)$'s  in \eqref{eq:basis detail}, $\mathcal{F}$ in \eqref{eq:TPS space}, and $J(f)$ in \eqref{eq:roughness},
and assume that $\mathrm{rank}(\bm{X})=d+1<n$. Then
\begin{enumerate}
\item $\mathcal{F}=\Big\{\displaystyle\sum_{k=1}^{n}a_k f_k(\cdot):a_k\in\mathbb{R}\Big\}$.
\item $\{f_{1}(\cdot),\ldots,f_{d+1}(\cdot)\}$
	is a basis of ${\displaystyle \left\{ g(\cdot)\in\mathcal{F}:J(g)=0\right\} }.$
\item For $k=d+2,\dots,n$, define
	\begin{equation}
\mathcal{F}_k=\Big\{g(\cdot)\in\mathcal{F}:\sum_{i=1}^{n}g(\bm{s}_{i})^{2}=1,\,
	\sum_{i=1}^{n}g(\bm{s}_{i})f_j(\bm{s}_{i})=0;\,j=1,\dots,k-1\Big\}.
	\label{eq:smoothest objective}
	\end{equation}
  Then $\displaystyle\mathop{\arg\min}_{g\in\mathcal{F}_k}J(g)=f_k(\cdot)$ and $J(f_k)=\lambda_{k-d-1}^{-1}$, 
  for $k=d+2,\dots,n$.
\end{enumerate}
\label{thm1}
\end{thm}

\begin{rem}
Let $\bm{f}_k=(f_k(\bm{s}_1),\dots,f_k(\bm{s}_n))'$; $k=1,\dots,n$. Then $\bm{f}'_{k}\bm{X}=\bm{0}$
and $\bm{f}'_k\bm{f}_{k^*}=I(k=k^{*})$, for $k,k^*=d+2,\dots,n$.
\end{rem}

\begin{figure}[tbh]\centering
\includegraphics[scale=0.55]{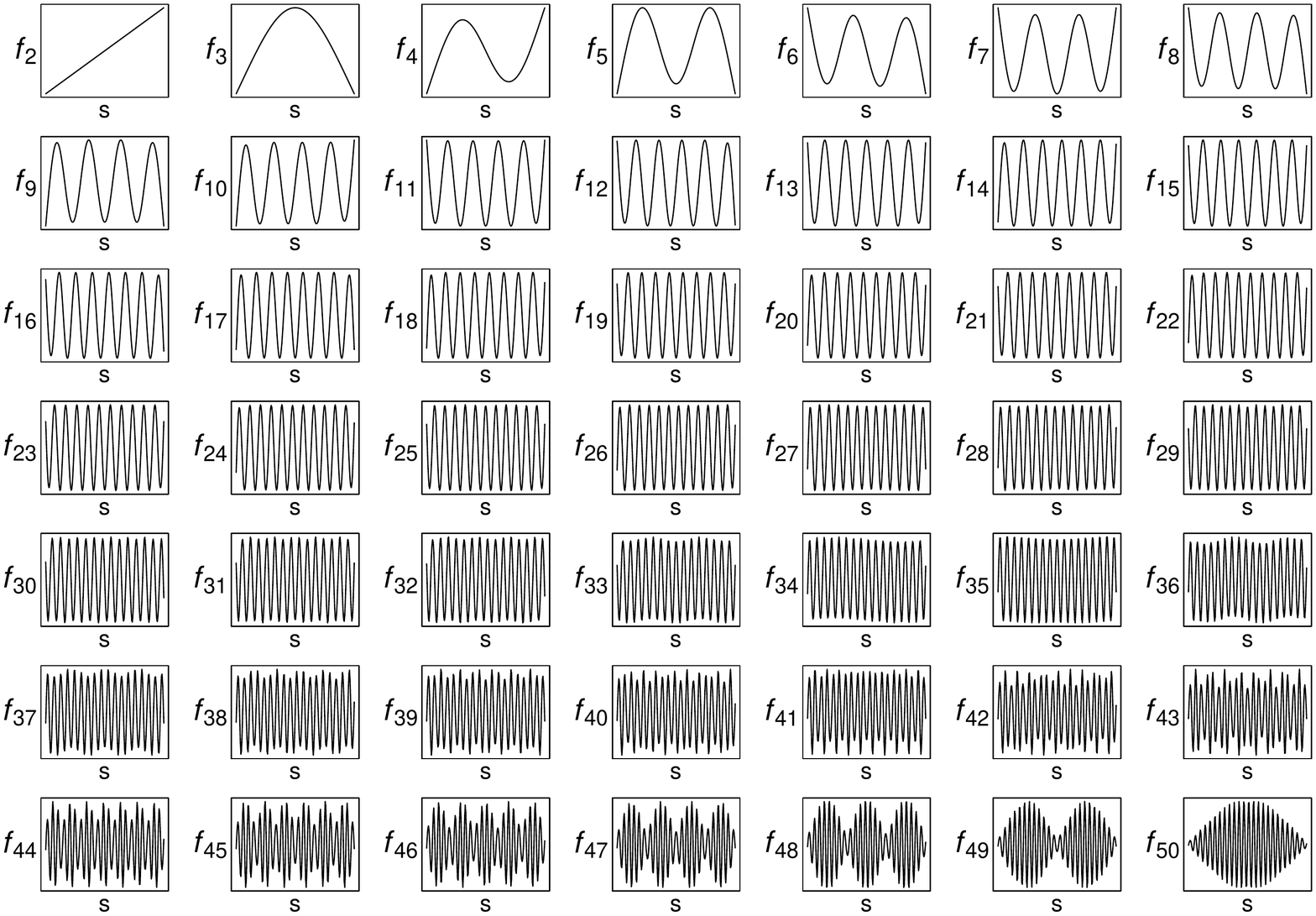}
\vspace{-0.5cm}
\caption{The proposed basis functions, $f_2(\cdot),\dots,f_{50}(\cdot)$.} 
\label{fig:nested resolution}
\end{figure}

\begin{rem}
The basis functions are given in a decreasing order in terms of their degrees of smoothness with
$0=J(f_1)=\cdots=J(f_{d+1})<J(f_{d+2})\leq\cdots\leq J(f_n)$. In addition, $f_k(\cdot)$ is the smoothest function that is orthogonal to $f_1(\cdot),\dots,f_{k-1}(\cdot)$, for $k=d+2,\dots,n$. This enables a spatial process to be more parsimoniously represented in the spatial
random-effects model, particularly when the underlying spatial covariance function is smooth.
A one-dimensional example of $f_2(\cdot),\dots,f_{50}(\cdot)$ with $n=50$ and $s_i=i/50$; $i=1,\dots,50$, is shown in Figure \ref{fig:nested resolution}. 
\end{rem}

\begin{rem}
Another basis of $\mathcal{F}$ is the Demmler-Reinsch basis \citep{demmler1975oscillation[TPS]} given by
\[
(h_1(\bm{s}),\dots,h_n(\bm{s}))'=\bm{U}'\big((\bm{X},\bm{\Phi}\bm{N})'(\bm{X},\bm{\Phi}\bm{N})\big)^{-1/2}\big(1,\bm{s}',\bm{\phi}(\bm{s})'\bm{N}\big)',
\]
where $\bm{N}$ is an $n\times(n-d-1)$ matrix such that $\bm{N}\bm{N}'=\bm{Q}$ and $ \bm{N}'\bm{N}=\bm{I}_{n-d-1}$,
and $\bm{U}\mathrm{diag}(a_1,\dots, a_n)\bm{U}'$ is the eigen-decomposition of
\[
\big((\bm{X},\bm{\Phi}\bm{N})'(\bm{X},\bm{\Phi}\bm{N})\big)^{-1/2}\left[
\begin{array}{cc}
\bm{0} & \bm{0}\\
\bm{0} & \bm{N}^{\prime}\bm{\Phi}\bm{N}
\end{array}
\right]\big((\bm{X},\bm{\Phi}\bm{N})'(\bm{X},\bm{\Phi}\bm{N})\big)^{-1/2},
\]
with $a_1\geq\cdots\geq a_n$. While $h_1(\cdot),\dots,h_n(\cdot)$
are orthogonal and satisfy $J(h_1)\leq \cdots\leq J(h_n)$, they generally do not have the property of
Theorem \ref{thm1} (iii). Additionally,
they are more expensive to compute since $\big((\bm{X},\bm{\Phi}\bm{N})'(\bm{X},\bm{\Phi}\bm{N})\big)^{-1/2}$
involves $O(n^{3})$  computations.
\end{rem}

Our method given by (\ref{eq:basis detail}) requires computing only the first $K$ eigenvalue and eigenvector pairs of
$\bm{Q}\bm{\Phi}\bm{Q}$ without the need to solve the full eigen-decomposition problem. In addition, we can
compute $\bm{Q}\bm{\Phi}\bm{Q}=\tilde{\bm{Q}}-\tilde{\bm{X}}'(\bm{X}^{\prime}\tilde{\bm{Q}})$ via $\tilde{\bm{X}}=(\bm{X}'\bm{X})^{-1}\bm{X}'$
and $\tilde{\bm{Q}}=\bm{\Phi}-(\bm{\Phi}\bm{X})\tilde{\bm{X}}$ to reduce the computations of $\bm{Q\Phi Q}$ from $O(n^{3})$ in terms of
direct matrix multiplication to  $O(n^2d)$. The first $K$ eigen-functions and eigenvalues can be efficiently obtained
using some numerical techniques, such as the QR method and the Lanczos method
(see e.g., \citealp{golub2000eigenvalue}; \citealt{ordonez2014pca}) via an R package such as ``bigpca"  or ``onlinePCA".
Both packages are available on Comprehensive R Archive Network (CRAN).

To know how the proposed basis functions perform in representing $C^{(0)}(\cdot,\cdot)$ of Example 1,
we consider six basis functions $f_1(\cdot),\dots,f_6(\cdot)$ (see Figure \ref{fig:motivation-part1}~(d1)) derived from
our method with the controlled points given at $s_i=i/50$; $i=1,\dots,50$, as in Figure \ref{fig:nested resolution}.
The best covariance function that minimizes \eqref{eq:ISE 1D} is shown in Figure \ref{fig:motivation-part1}~(d2).
Clearly, it provides a much better approximation to the true spatial covariance function than
those in Figure \ref{fig:motivation-part1}~(b2) and (c2)
based on $\bm{f}^{(1)}(\cdot)$ and $\bm{f}^{(2)}(\cdot)$.

\begin{figure}[bt]\centering
\begin{tabular}{cc}
\includegraphics[scale=0.18]{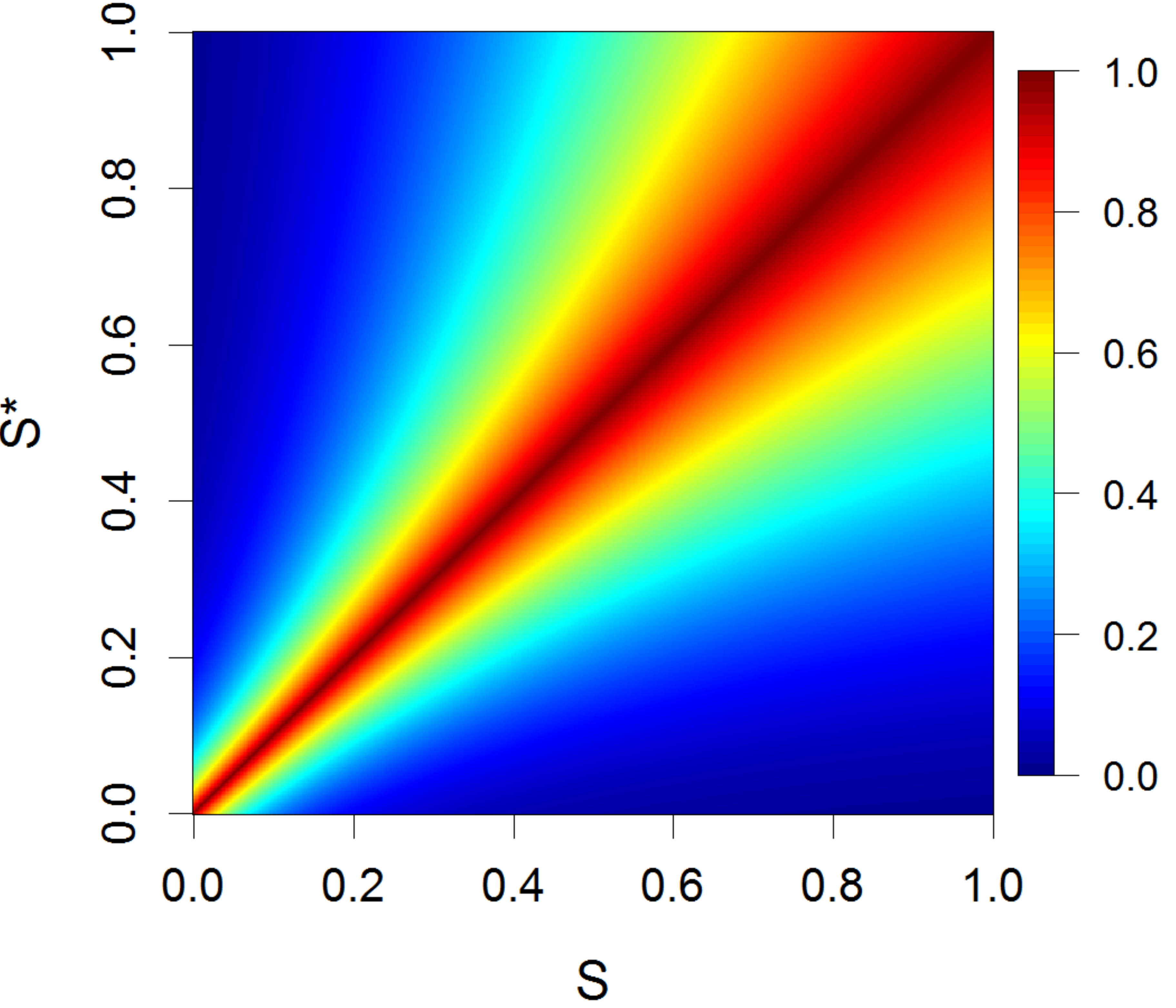}  & \includegraphics[scale=0.18]{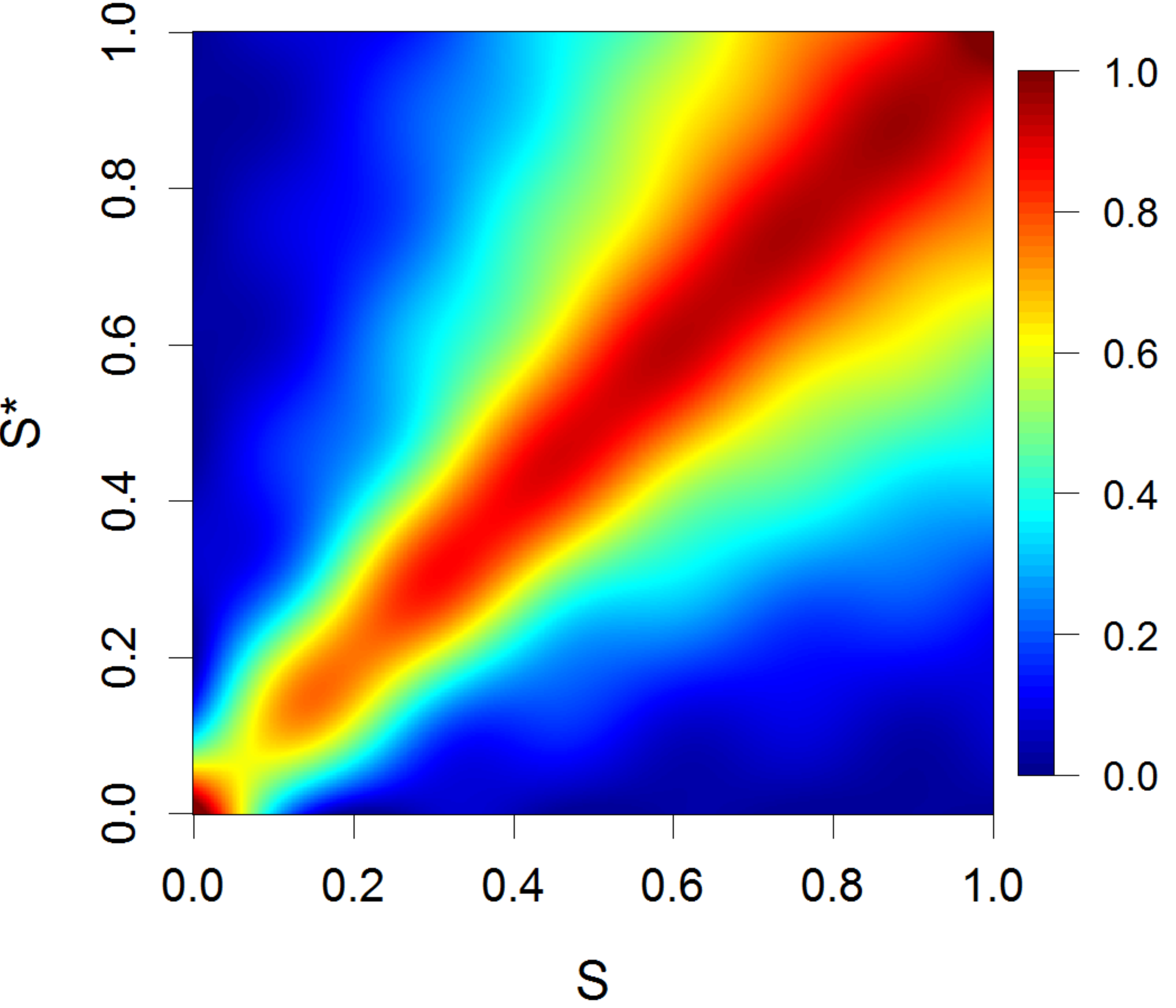} \vspace{0cm}\\		
(a)  & (b) \medskip
\tabularnewline
\includegraphics[scale=0.18]{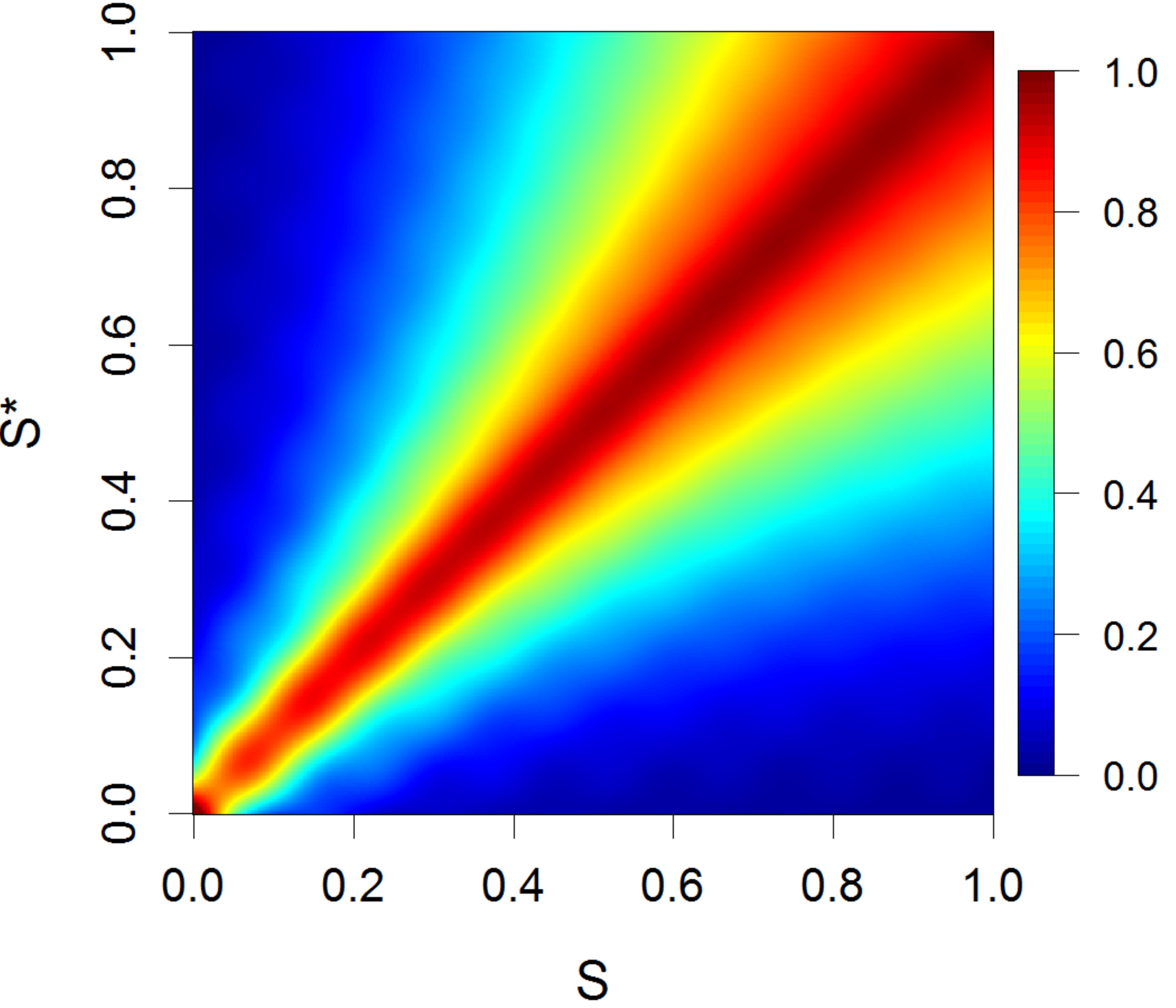}  &  \includegraphics[scale=0.18]{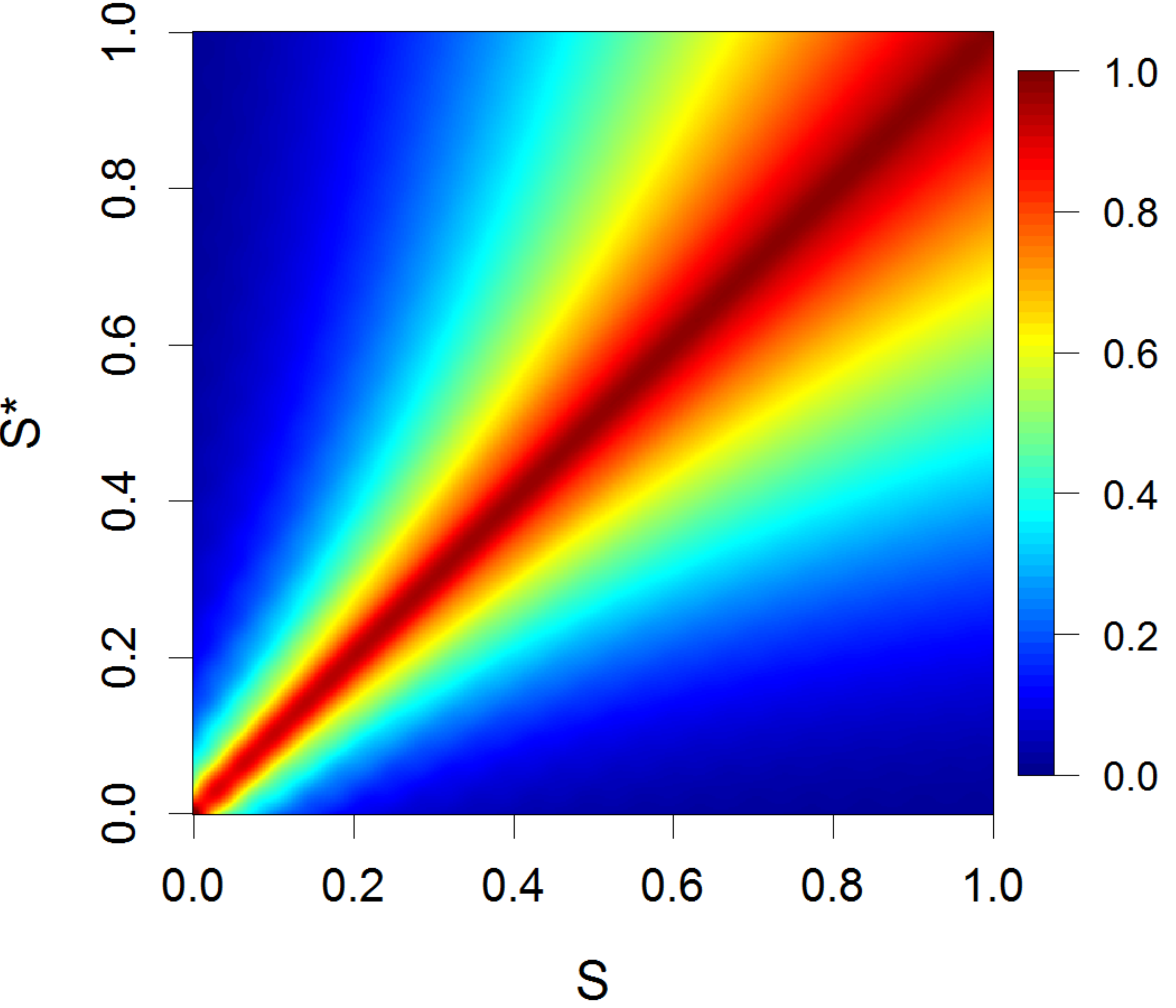}\vspace{0cm}\\
(c) & (d)\tabularnewline
\end{tabular}
\caption{(a) A nonstationary spatial covariance function; (b) covariance function approximation based on $8$ basis functions;
(c) covariance function approximation based on $15$ basis functions;  (d) covariance function approximation based on $30$ basis functions.}
\label{fig:multi-resolution example} 
\end{figure}

To illustrate how the proposed basis functions provide a multi-resolution covariance function representation,
we consider a spatially deformed exponential covariance function: 
\[
C(s,s^*)=\exp \big\{-2 \big|(s+0.5)^{-1.5}-(s^{*}+0.5)^{-1.5}\big|\big\};\quad s,s^*\in[0,1]
\] 
(see Figure \ref{fig:multi-resolution example}~(a)), which is a nonstationary covariance function
constructed by applying a deformation transformation ($s\rightarrow(s+0.5)^{-1.5}$) to a stationary covariance function
as in \citet{sampson1992nonparametric[deformation]}.
We apply our basis functions (see Figure \ref{fig:nested resolution})
to approximate this covariance function, where the controlled points are given at $s_i=i/50$; $i=1,\dots,50$.
The results for three different numbers of basis functions ($K=8,15,30$) are shown in Figure
\ref{fig:multi-resolution example}~(b)-(d), respectively.
As you can see, large-scale features can be captured even if $K$ is merely $8$. On the other hand,
finer-resolution details are captured by $f_k(\cdot)$ with larger $k$ values.
 
The proposed class of basis functions is even more effective in the two-dimensional space.
Suppose that we would like to approximate an exponential covariance function,
$C(\bm{s},\bm{s}^*)=20\exp(-0.4\|\bm{s}-\bm{s}^*\|)$ for
$\bm{s},\bm{s}^*\in[0,1]^2$, using $\bm{f}(\bm{s})'\bm{M}\bm{f}(\bm{s})$. We compare between a conventional method
and our method. For a conventional method, we consider the natural TPS functions for $\bm{f}(\cdot)$ formed by $1$, $x_1$, $x_2$ and
\[
\frac{1}{8\pi}\bigg\| \bm{s}-\bigg(\frac{\ell_1}{L+1},\frac{\ell_2}{L+1}\bigg)'\bigg\|^{2}
\log\bigg\{\bigg\|\bm{s}-\bigg(\frac{\ell_1}{L+1},\frac{\ell_2}{L+1}\bigg)'\bigg\|\bigg\};\quad 1\leq \ell_1,\ell_2\leq L,
\]
with their centers regularly location in $[0,1]^2$ for $L\in\{3,5,7,9,11,13\}$,
corresponding to a total of $\{12,28,52,84,124,172\}$ basis functions.
We apply our method with the control points, $\{((2j_1-1)/36,(2j_2-1)/36):1\leq j_1,j_2\leq 18\}$, regularly located in
$[0,1]^2$, and consider the same numbers of basis functions for comparison.
The performance between the conventional basis functions and the proposed basis functions is shown in Table \ref{tab:motivation2D}.
For all cases, the proposed basis functions provide much better approximation ability than the conventional basis functions. 

\begin{table}
\caption{ISE performance between TPS basis functions and the proposed basis functions for various numbers of functions.\label{tab:motivation2D}}
\medskip
\begin{center}
\begin{tabular}{ccc}\hline
number of basis functions  & TPS & Proposed\tabularnewline\hline 
$3^{2}$+3 & 0.09462 & 0.01895\tabularnewline
$5^{2}$+3 & 0.01505 & 0.00301\tabularnewline
$7^{2}$+3 & 0.00416 & 0.00085\tabularnewline
$9^{2}$+3 & 0.00155 & 0.00037\tabularnewline
$11^{2}$+3 & 0.00070 & 0.00021\tabularnewline
$13^{2}$+3 & 0.00037 & 0.00015\tabularnewline\hline 
\end{tabular}
\end{center}
\end{table}

\section{Parameter Estimation} \label{sec: Estimation and Selection}

Consider the spatial random-effects model given by (\ref{eq:uni-obs-equation-z})
and (\ref{eq:uni-general-form-y}). For simplicity, we assume that $\mu(\bm{s},t)=0$ and
$\sigma_\epsilon^2$ is known, since $\sigma_\epsilon^2$ and $\sigma_\xi^2$ are confounded together
unless some additional information is available.
Given the basis functions $f_1(\cdot),\dots,f_K(\cdot)$,
the parameters that need to be estimated are $\bm{M}$, which has to be non-negative definite, and $\sigma_\xi^2\geq 0$. Although the ML estimates
$\hat{\bm{M}}_{K}$ and $\hat{\sigma}_{\xi,K}^2$ of $\bm{M}$ and $\sigma_\xi^2$ can be computed using the EM algorithm \citep{katzfuss2009EM[clm]},
as shown in the following theorem, a closed-form expression for $\hat{\bm{M}}_{K}$ can be derived with its proof given in Appendix.
The estimate $\hat{\sigma}_{\xi,K}^2$ can be computed using a simple one-dimensional optimization method.

\begin{thm}
Consider the model given by (\ref{eq:uni-obs-equation-z}) and (\ref{eq:uni-general-form-y}) with $\mu(\bm{s},t)=0$ and
$\sigma_\epsilon^2$ known. Then the ML estimates of $\bm{M}$ and $\sigma_{\xi}^2$ are given by
\begin{align*}
	\hat{\sigma}_{\xi,K}^2
=&~ \mathop{\arg\min}_{\sigma_\xi^2}\left[\frac{\mathrm{tr}(\bm{S})}{\sigma_\xi^2+\sigma_\epsilon^2}+
	\sum_{k=1}^K\bigg\{\log\big(\hat{d}_{K,k}+\sigma_\xi^2+\sigma_\epsilon^2\big)-
	\frac{d_{K,k}\hat{d}_{K,k}}{\sigma_\xi^2+\sigma_\epsilon^{2}}\bigg\}
	+(n-K)\log(\sigma_\xi^2+\sigma_\epsilon^2)\right],\\
	\hat{\bm{M}}_K
=&~ \left(\bm{F}'_{K}\bm{F}_{K}\right)^{-1/2}\bm{P}_K\,\mathrm{diag}\big(\hat{d}_{K,1},\dots,\hat{d}_{K,K}\big)
	\bm{P}'_K\left(\bm{F}'_{K}\bm{F}_{K}\right)^{-1},
\end{align*}
where $\bm{S}=\displaystyle\sum_{t=1}^{T}\bm{z}_{t}\bm{z}'_{t}/T$, $\bm{F}_K=(\bm{f}_{1},\dots,\bm{f}_K)$,
$\bm{f}_k=(f_k(\bm{s}_1),\dots,f_k(\bm{s}_n))'$;\quad $k=1,\dots,K$, $\bm{P}_K\,\mathrm{diag}(d_{K,1},\dots,d_{K,K})\bm{P}'_K$
is the eigen-decomposition of  $(\bm{F}'_K\bm{F}_K)^{-1/2}\bm{F}'_{K}\bm{S}\bm{F}_K\left(\bm{F}'_K\bm{F}_K\right)^{-1/2}$,
and $\hat{d}_{K,k}=\max\left(d_{K,k}-\hat{\sigma}_{\xi,K}^2-\sigma_{\epsilon}^2,0\right)$; $k=1,\dots,K$.
\label{thm2}
\end{thm}

In practice, we propose to select $K\in\{d+1,\dots,K^*\}$ for a sufficiently large $K^*$
using Akaike's information criterion (AIC, \citealp{akaike1973,akaike1974}):
\begin{align*}
	\mathrm{AIC}(K)
=&~ T\log\big|\hat{\bm{\Sigma}}_k\big|+T\mathrm{tr}\big(\bm{S}\hat{\bm{\Sigma}}^{-1}_K\big)+K^2+K+2\\
=&~ \frac{T\mathrm{tr}(\bm{S})}{\hat{\sigma}_{\xi,K}^2+\sigma_\epsilon^2}+T\sum_{k=1}^K\bigg\{
	\log\big(\hat{d}_{K,k}+\hat{\sigma}_{\xi,K}^2
	+\sigma_\epsilon^2\big)-\frac{d_{K,k}\hat{d}_{K,k}}{\hat{\sigma}_{\xi,K}^{2}+\sigma_\epsilon^2}\bigg\}+K^2+K+2,
\end{align*}

\noindent where $\hat{\bm{\Sigma}}_K=\bm{F}_{K}\hat{\bm{M}}_K\bm{F}'_{K}+(\hat{\sigma}_{\xi,K}^2+\sigma_\epsilon^2)\bm{I}_n$.
Then the final number of basis functions selected by AIC is $\hat{K}=\displaystyle\mathop{\arg\min}_{d+1\leq K\leq K^*}\mathrm{AIC}(K)$.
Plugging in $\hat{\bm{M}}_{\hat{K}}$ and $\hat{\sigma}_{\xi,\hat{K}}^2$ into
the best linear unbiased predictor of $y(\bm{s},t)$, we obtain
\begin{equation}
\hat{y}(\bm{s},t)=\big\{\bm{f}(\bm{s})'\bm{\hat{M}}_{\hat{K}}\bm{F}'_K+\hat{\sigma}_{\xi,\hat{K}}^2(I(\bm{s}=\bm{s}_{1}),\dots,I(\bm{s}=\bm{s}_n))
\big\}\bm{\hat{\Sigma}}^{-}_{\hat{K}}\bm{z}_t\:;\quad \bm{s}\in D,\,t=1,\dots,T,
\label{eq:yhat}
\end{equation}
where $\bm{\hat{\Sigma}}^{-}_{\hat{K}}$ is the Moore-Penrose inverse of $\bm{\hat{\Sigma}}_{\hat{K}}$
and can be efficiently computed by
\begin{small}
\begin{equation}
\begin{cases}
\displaystyle\frac{1}{\hat{\sigma}_{\xi,\hat{K}}^2+\sigma_\epsilon^2}\left\{\bm{I}_n-\bm{L}_{\hat{K}}\bm{P}_{\hat{K}}\,\mathrm{diag}\bigg(
	\frac{d_{\hat{K},1}}{d_{\hat{K},1}+\hat{\sigma}_{\xi,\hat{K}}^2+\sigma_\epsilon^2},\dots,
	\frac{d_{\hat{K},\hat{K}}}{d_{\hat{K},\hat{K}}+\hat{\sigma}_{\xi,\hat{K}}^2+\sigma_\epsilon^2}\bigg)
	\bm{P}'_{\hat{K}}\bm{L}'_{\hat{K}}\right\}; & \mathrm{if}~\hat{\sigma}_{\xi,\hat{K}}^2+\sigma_\epsilon^2>0,\\
\bm{L}_{\hat{K}}\bm{P}_{\hat{K}}\,\big\{\mathrm{diag}\big(d_{\hat{K},1},\dots,d_{\hat{K},\hat{K}}\big)\big\}^{-}\bm{P}'_{\hat{K}}
	\bm{L}'_{\hat{K}}; & \mathrm{if}~\hat{\sigma}_{\xi,\hat{K}}^2=\sigma_\epsilon^2=0,
\end{cases}
\label{eq:fast inverse}
\end{equation}
\end{small}
and $\bm{L}_{\hat{K}}=\bm{F}_{\hat{K}}(\bm{F}'_{\hat{K}}\bm{F}_{\hat{K}})^{-1/2}$.

\section{Numeric Examples}

\subsection{Simulation}

\begin{figure}[tb]\centering
\begin{tabular}{cc}
\includegraphics[scale=0.18]{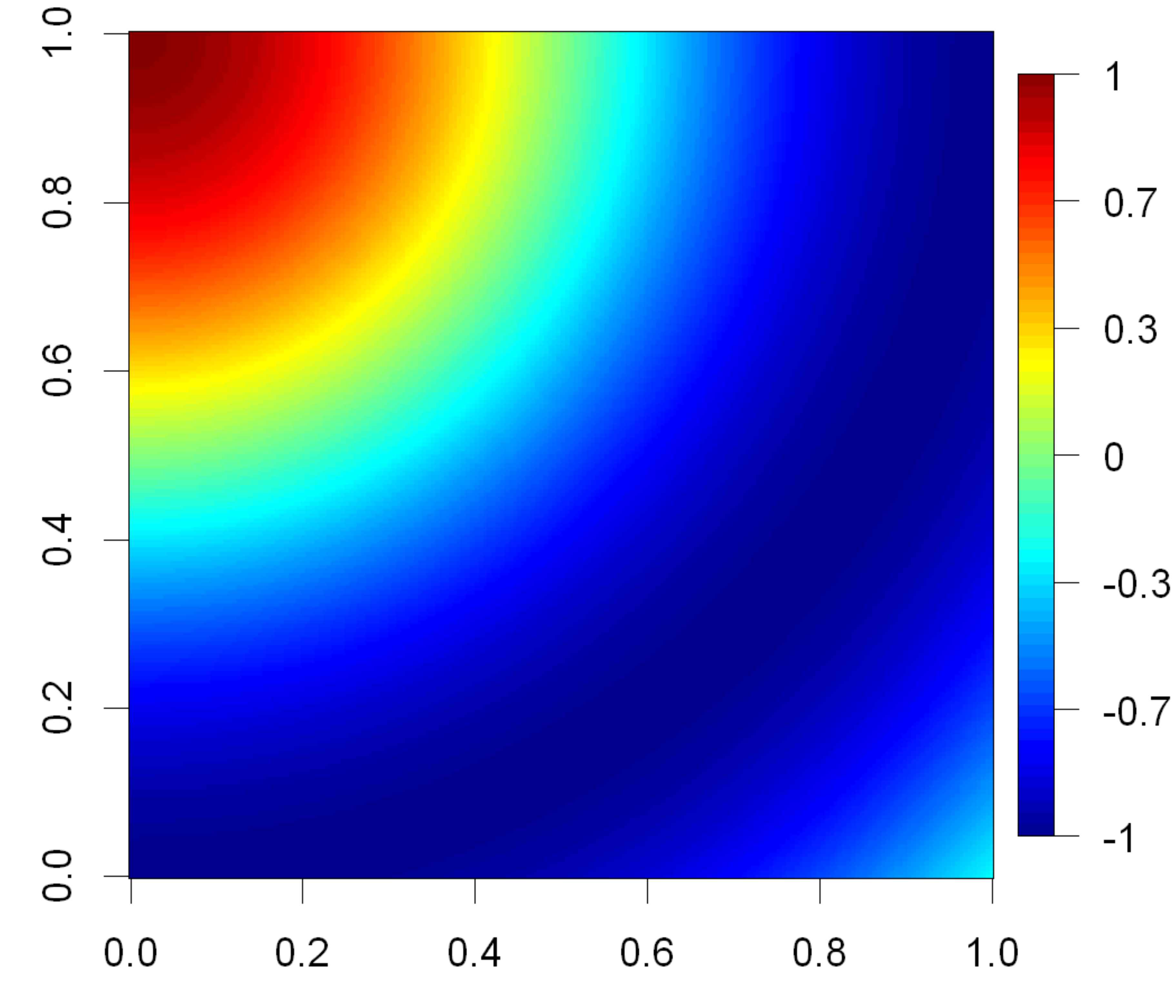} & \includegraphics[scale=0.18]{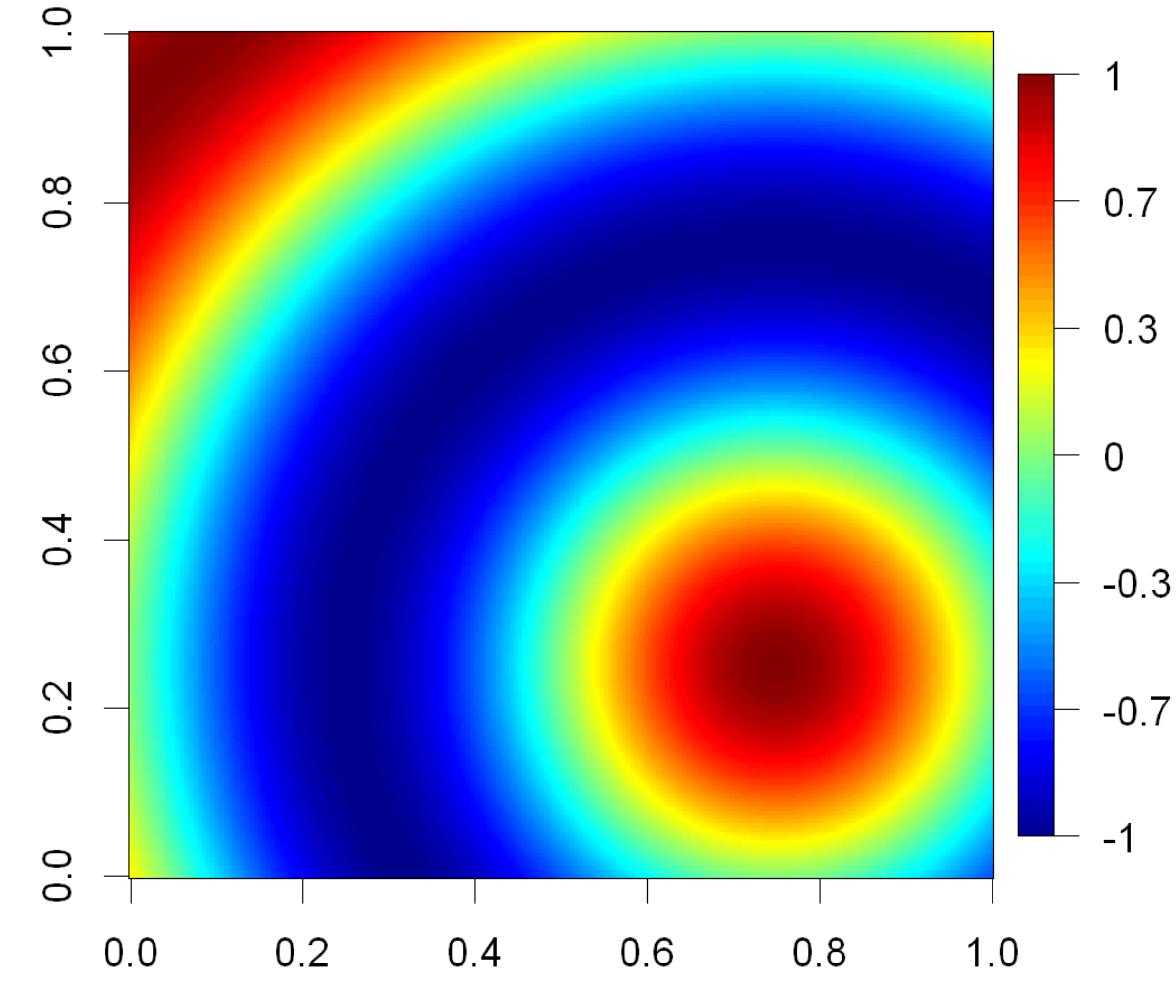}\tabularnewline
(a)  & (b)\tabularnewline
\end{tabular}
\caption{Basis functions in a spatial random-effects model: (a) $f_1(\cdot)$; (b) $f_2(\cdot)$.}	
\label{fig:2D} 
\end{figure}

In the simulation, we considered spatial processes, $\{y(\bm{s},t):\bm{s}\in[0,1]^2\}$ for $t=1,\dots,50$, generated from
\eqref{eq:uni-general-form-y} with $\mu(\bm{s},t)=0$, $f_1(\bm{s})=\cos(\pi\|\bm{s}-(0,1)'\|)$, $f_2(\bm{s})=\cos(2\pi\|\bm{s}-(3/4,1/4)'\|)$,
and $(w_1(t),w_2(t))'\sim N(\bm{0},\mathrm{diag}(25,9))$, where $f_1(\cdot)$ and $f_2(\cdot)$
are shown in Figure \ref{fig:2D}. We generated data, $\bm{z}_1,\dots,\bm{z}_{50}$, according to
\eqref{eq:uni-obs-equation-z} with $n=100$ and $\sigma_\epsilon^2=3$,
where $\bm{s}_{1},\dots,\bm{s}_{n}$ were taken from $D=[0,1]^{2}$ using simple random sampling.

We applied the spatial random-effects model of \eqref{eq:uni-obs-equation-z} and \eqref{eq:uni-general-form-y} 
and the ML estimates given by Theorem \ref{thm2} to estimate the underlying spatial covariance function
with $\sigma_\epsilon^2=3$ assumed known. We considered commonly used
bisquare basis functions given in \eqref{eq:1D example} with six different layouts for
function centers and radii at two resolutions (see Table \ref{tab:FRK-layout}).
We applied the proposed basis functions and selected the number of basis functions among $K\in\{3,\dots,20\}$ using AIC.
We also considered the exponential covariance model and the true covariance function for comparison.
All the model parameters were estimated by ML.

\begin{table}\centering
\caption{Various layouts for centers of the bisque basis functions.\label{tab:FRK-layout}}
\medskip
\begin{tabular}{c|cc|cc|c}
\hline 
Layout & \multicolumn{2}{|c}{Coarse Resolution} & \multicolumn{2}{|c|}{Fine Resolution} & K\tabularnewline
\cline{2-5} & Center & Radius & Center & Radius\\\hline
1 & $\{0,1\}^2$ & $3/2$ & $\{1/4,3/4\}^2$ & $3/4$ & $8$\\
2 & $\{1/6,5/6\}^2$ & $1$ & $\{0,1/2,1\}^2$ & $3/2$ & $13$\\
3 & $\{1/6,5/6\}^2\cup(1/2,1/2)$ & $\sqrt{2}/2$ & $\{0,1/2,1\}^2$ & $3/2$ & $14$\\
4 & $\{0,1/2,1\}^2$ & $3/4$ & $\{1/6,1/2,5/6\}^2$ & $1/2$ & $18$\\
5 & $\{1/6,5/6\}^2$ & $1$ & $\{0,1/3,2/3,1\}^2$ & $1/2$ & $20$\\
6 & $\{1/6,5/6\}^2\cup(1/2,1/2)$ & $\sqrt{2}/2$ & $\{0,1/3,2/3,1\}^2$ & $1/2$ & $21$\\\hline 
\end{tabular}

\end{table}

The performance of various methods was compared in terms of the mean-squared-prediction-error
(MSPE) criterion:
\[
\frac{1}{50}\sum_{t=1}^{50}\int_{[0,1]^2}E(\hat{y}(\bm{s},t)-y(\bm{s},t))^{2},
\]
where $\hat{y}(\bm{s},t)$ is a generic predictor of $y(\bm{s},t)$ obtained from simple kriging based on $\bm{z}_t$
using an (estimated) spatial covariance model.
The results based on 200 simulation replicates are shown in Table \ref{tab:FRK-MSPE}.
Not surprisingly, bisquare basis functions perform well for some cases but poorly for others.
In contrast, our method performs better than all the other spatial covariance estimation
methods by having a smaller averaged MSPE value.
The first and the third quantiles for the distribution of the number of basis functions
selected by AIC are about 10 and 12, indicating that only a small number of basis functions is required.

\begin{table}[tb]\centering
\caption{Averaged MSPEs for various methods based on 200 simulation replicates. Values given in parentheses are the corresponding standard errors.\label{tab:FRK-MSPE}}
\medskip
\begin{tabular}{ccccccccc}
\hline 
True & Exponential & Our & \multicolumn{6}{c}{Bisque Basis Functions}\tabularnewline
\cline{4-9} &&& 1 & 2 & 3 & 4 & 5 & 6\\\hline
0.123 & 1.234 & 0.646 & 0.694 & 0.872 & 1.063 &  0.962 & 1.013 & 1.191\tabularnewline
(0.015) & (0.017) & (0.015) & (0.024) & (0.018) & (0.031) & (0.034) & (0.032) & (0.037)\tabularnewline
\hline 
\end{tabular}

\end{table}

\subsection{Application to Canadian Temperature Data}

We applied the proposed method to an average daily temperature dataset. The data, available in the ``fda"
package on CRAN, consist of average temperatures for each day of the year at 35 weather stations in Canada, which are
averaged over years 1960 to 1994.
They have been analyzed by \citet{ramsay1991some} and \citet{silverman2005functional}
using functional data analysis techniques without considering spatial dependence.

Let $z(\bm{s}_i,t)$ be the average daily temperature at location $\bm{s}_i$ and day $t$,
where $\bm{s}_{i}$ is given with coordinates in latitude and longitude in units of degrees.
We considered the spatial random-effects model of (\ref{eq:uni-obs-equation-z}) and (\ref{eq:uni-general-form-y})
with $n=35$ and $T=365$.
Since the temporal patterns are known to be different at different stations (see e.g., \citealp{silverman1995incorporating}),
we considered a semiparametric model (\citealp{buja1989linear}) for $\mu(\bm{s},t)$ with station-specific quadratic effects:
\begin{equation}
\mu(\bm{s},t)=m_{0}(t)+m(\bm{s})+\ell(\bm{s}) t+q(\bm{s}) t^2;\quad \bm{s}\in D,\, t=1,\dots,365,
\label{eq:Canada mean function}
\end{equation}
where $m_0(\cdot),m(\cdot),\ell(\cdot)$ and $q(\cdot)$ are unknown smooth functions, and for identification purpose,
we assume $\displaystyle\sum_{i=1}^{35}m(\bm{s}_i)=\displaystyle\sum_{i=1}^{35}\ell(\bm{s}_i)=\displaystyle\sum_{i=1}^{35}q(\bm{s}_i)=0$.

\begin{figure}[bt]\centering
\begin{tabular}{cc}
\includegraphics[scale=0.32]{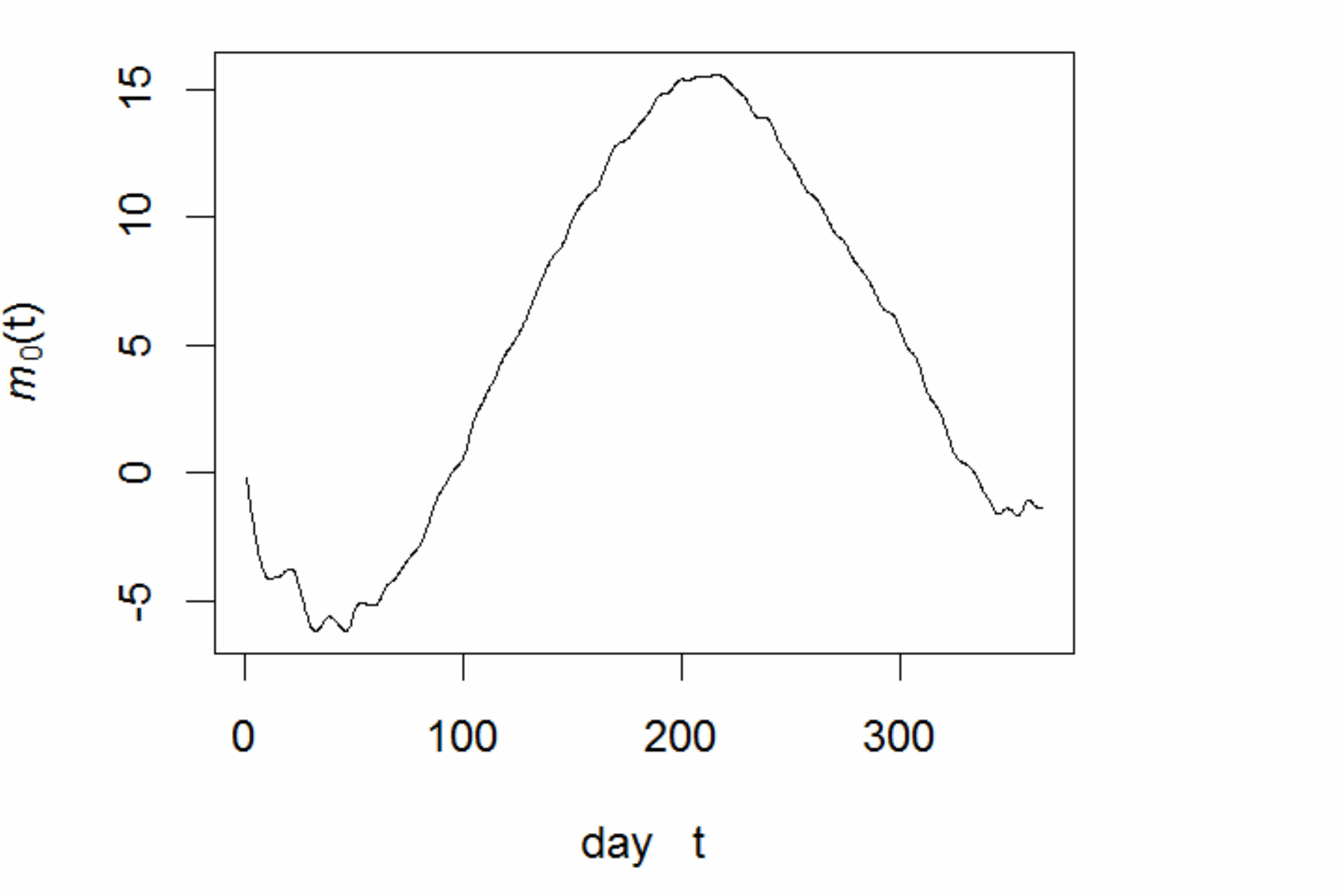}  & \includegraphics[scale=0.32]{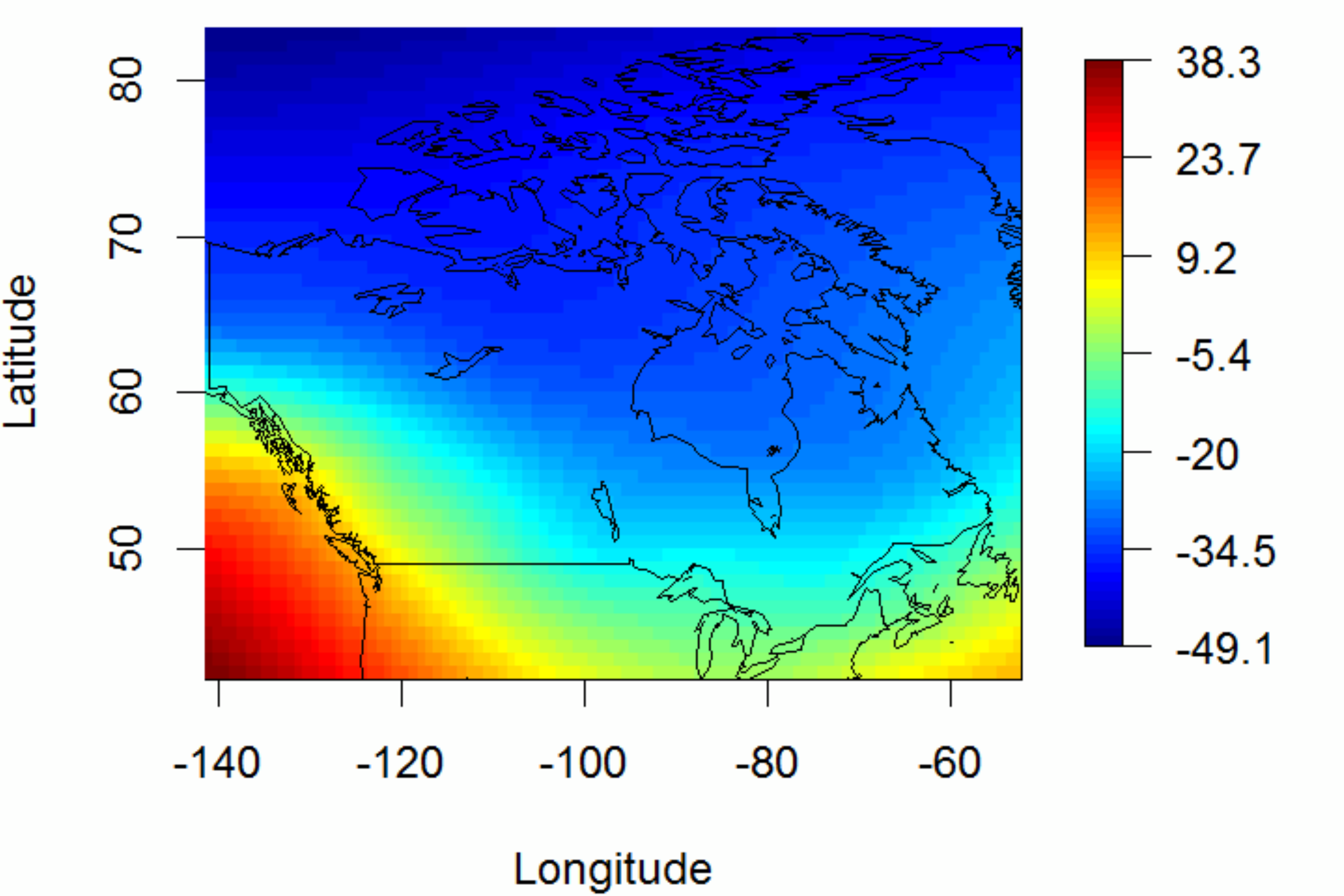} \vspace{-0.2cm}\\
(a)  & (b)\tabularnewline
\includegraphics[scale=0.32]{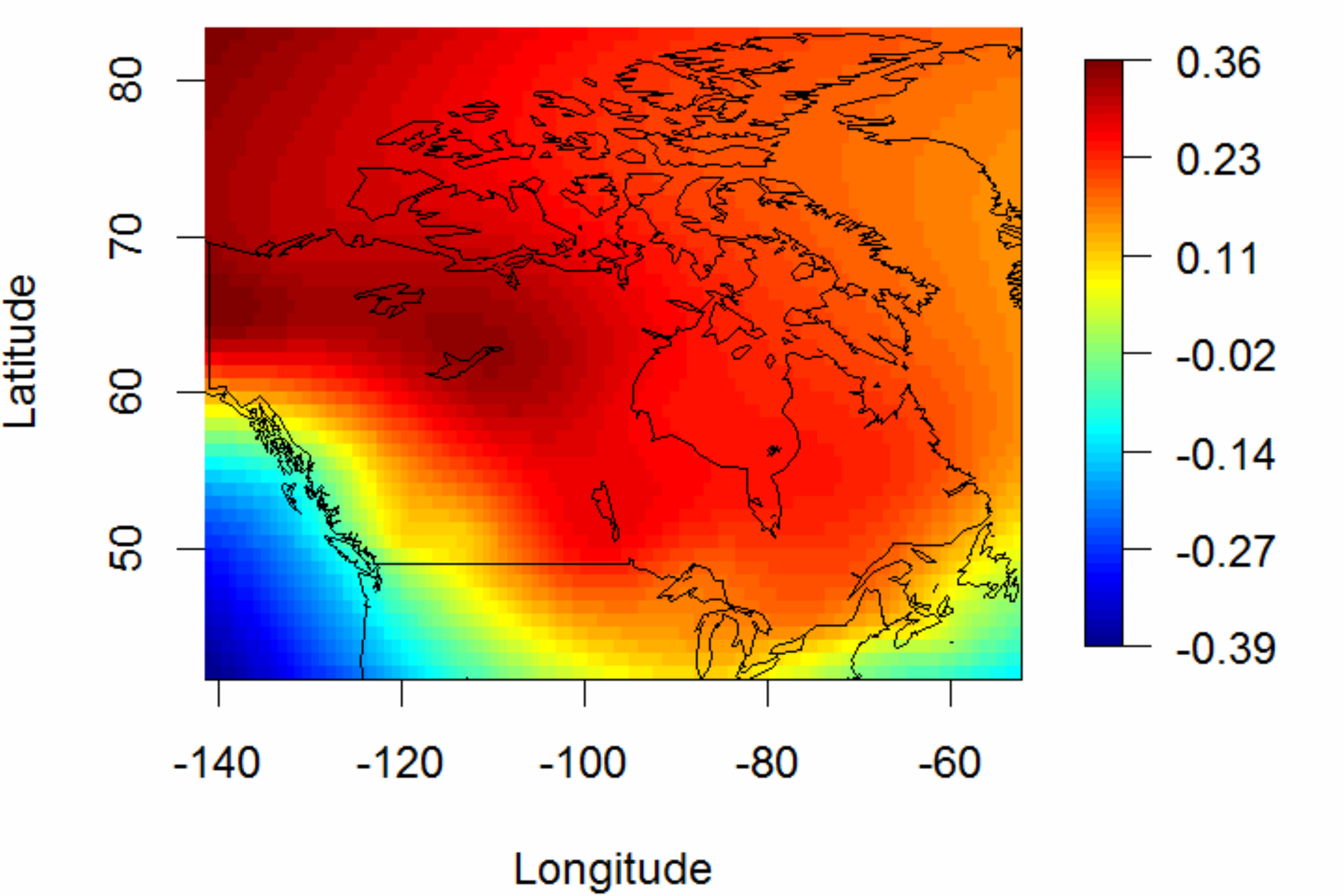}  & \includegraphics[scale=0.32]{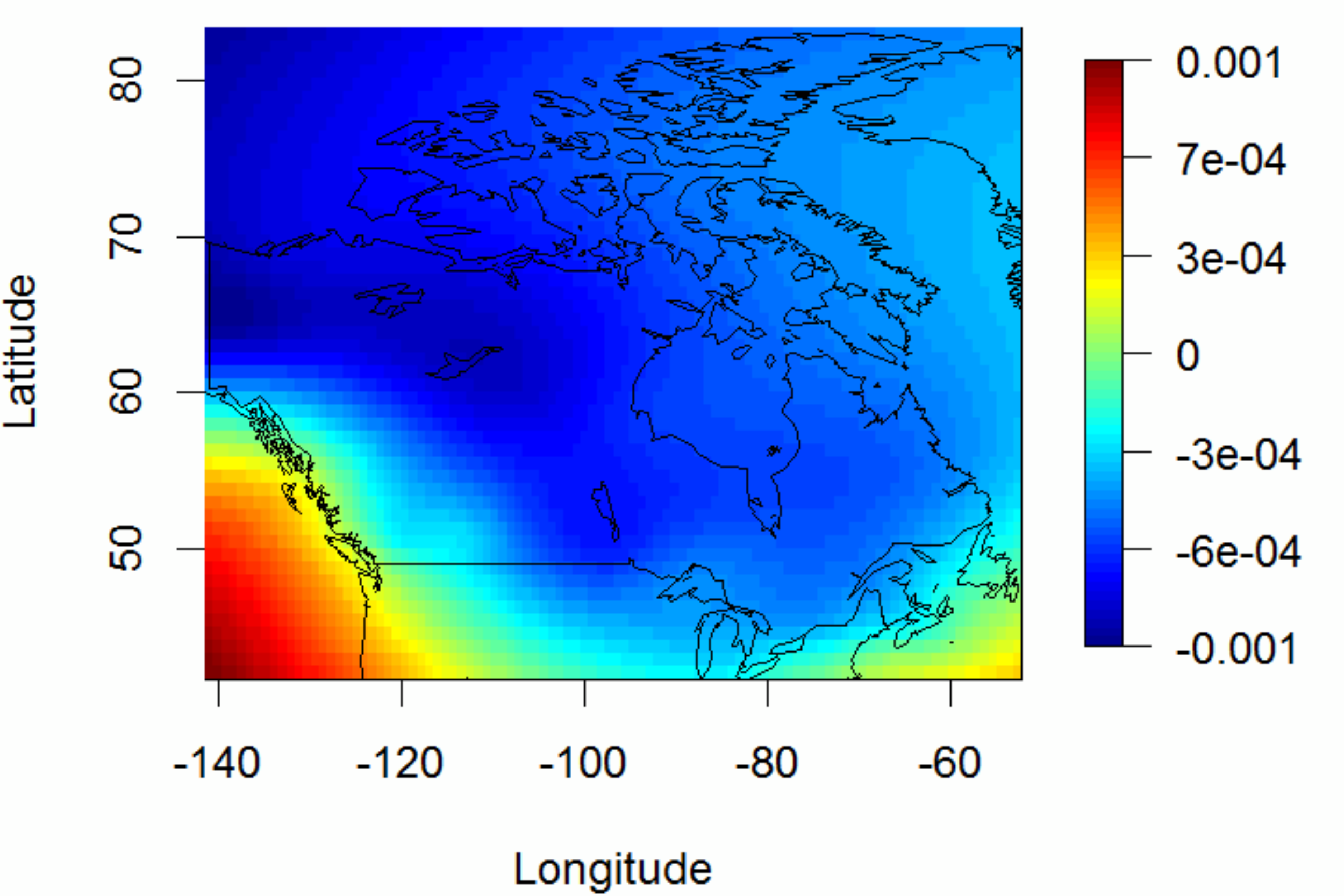}\vspace{-0.1cm}\\
(c) & (d)\tabularnewline
\end{tabular}
\caption{Estimated functions in \eqref{eq:Canada mean function}:
(a) $\hat{m}_{0}(t)$; (b) $\hat{m}(\bm{s})$; (c) $\hat{\ell}(\bm{s})$; (d) $\hat{q}(\bm{s})$.}
\label{fig:coefficient maps} 
\end{figure}

We considered a two-step procedure to fit $\mu(\cdot,\cdot)$
with the smoothness parameter selected by using Mallow's $C_p$ (\citealp{hastie1990generalized}).
First, we obtained the estimates $\hat{m}_i$, $\hat{\ell}_i$ and $\hat{q}_i$ of
$m(\bm{s}_i)$, $\ell(\bm{s}_i)$ and $q(\bm{s}_i)$ for $i=1,\dots,35$, and the estimate $\hat{m}_0(\cdot)$ of $m_0(\cdot)$
using the R package ``gam" available on CRAN (see Figure~\ref{fig:coefficient maps}~(a)).
Then we separately applied smoothing splines to $\hat{m}_i$'s, $\hat{\ell}_i$'s and $\hat{q}_i$'s and obtained the estimates
$\hat{m}_i(\cdot)$, $\hat{\ell}(\cdot)$ and $\hat{q}(\cdot)$ (see Figure~\ref{fig:coefficient maps}~(b)-(d)) with the smoothing parameter selected by
generalized cross-validation (\citealp{golub1979generalized}). Then we assume that $\mu(\bm{s},t)$ is known as
$\hat{\mu}(\bm{s},t)=\hat{m}_{0}(t)+\hat{m}(\bm{s})+\hat{\ell}(\bm{s})t+\hat{q}(\bm{s}) t^2$ for covariance function estimation.

We randomly divided the data into two parts with one part consisting of 185 time points as the training data, and the other part
consisting of 180 time points as the testing data.
We applied the spatial random-effects model of \eqref{eq:uni-obs-equation-z} and \eqref{eq:uni-general-form-y}.
We assumed that $\sigma_\xi^2=0$, but $\sigma_\epsilon^2$ is unknown, and applied ML with the proposed basis functions to estimate the underlying spatial covariance function.
We also considered applying the exponential covariance model to estimate covariance function with the parameters estimated
by ML.

\begin{figure}\centering
\begin{tabular}{cc}
\includegraphics[scale=0.32]{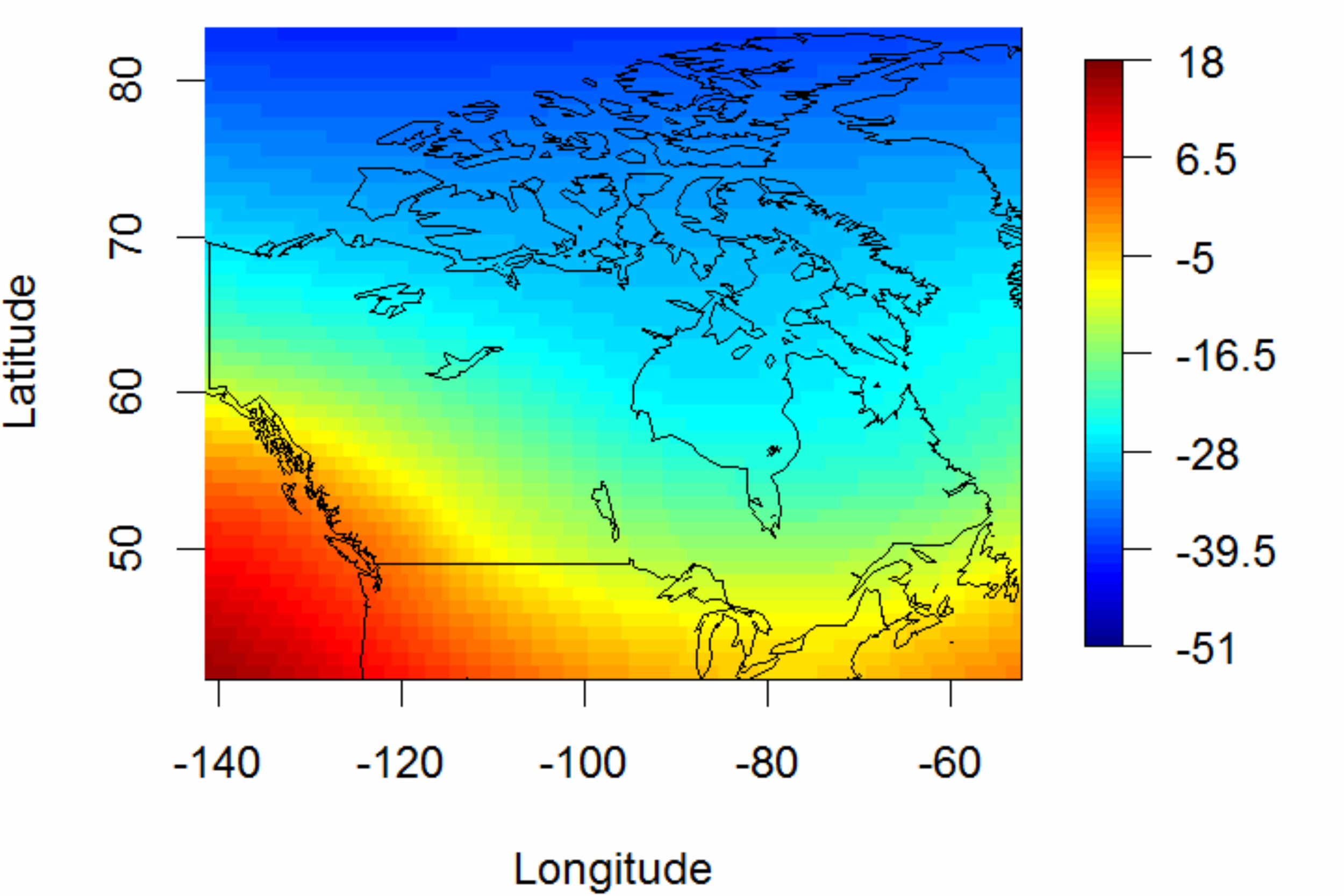}  & \includegraphics[scale=0.32]{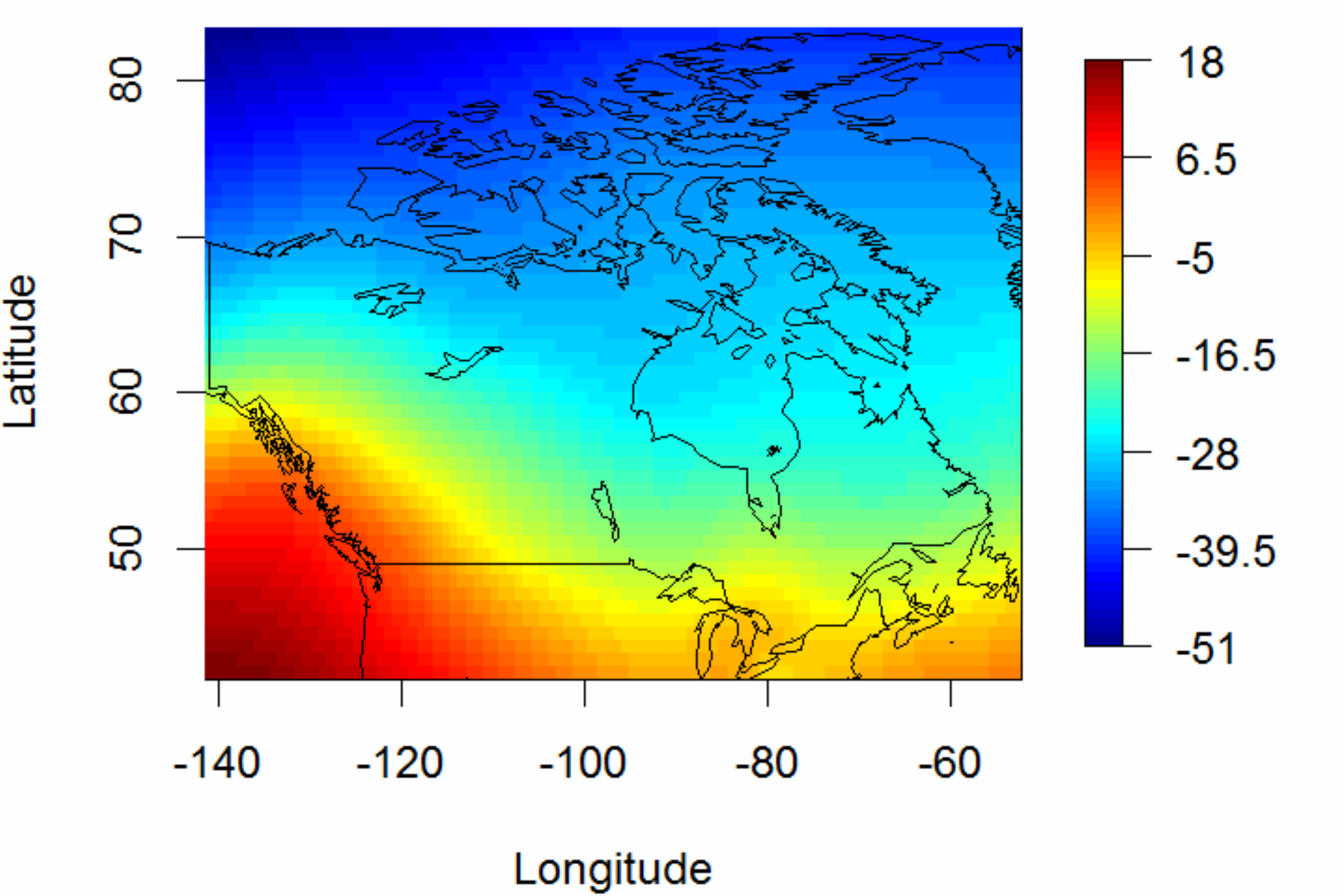} \tabularnewline
(a1)  & (a2)\tabularnewline
\includegraphics[scale=0.32]{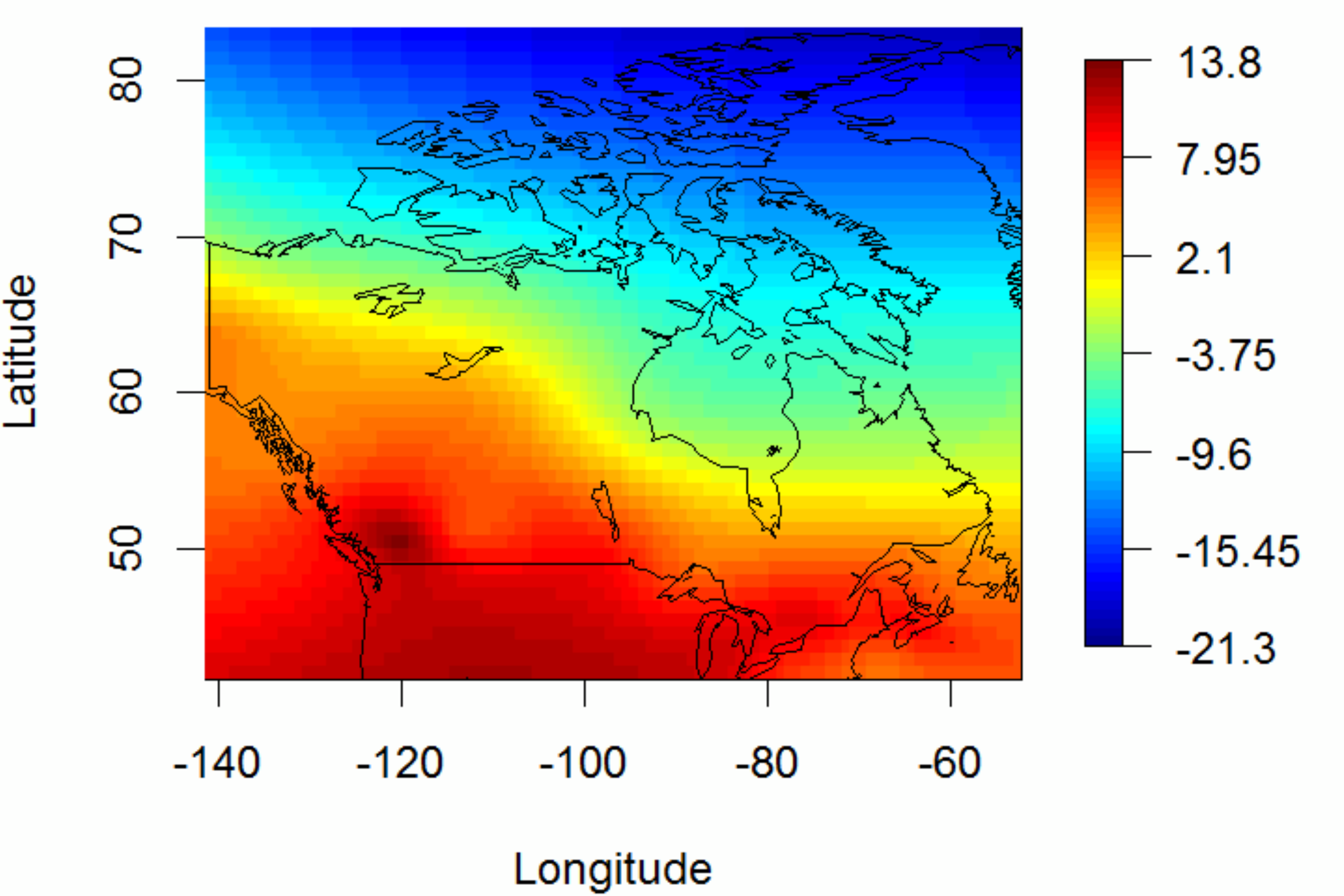}  & \includegraphics[scale=0.32]{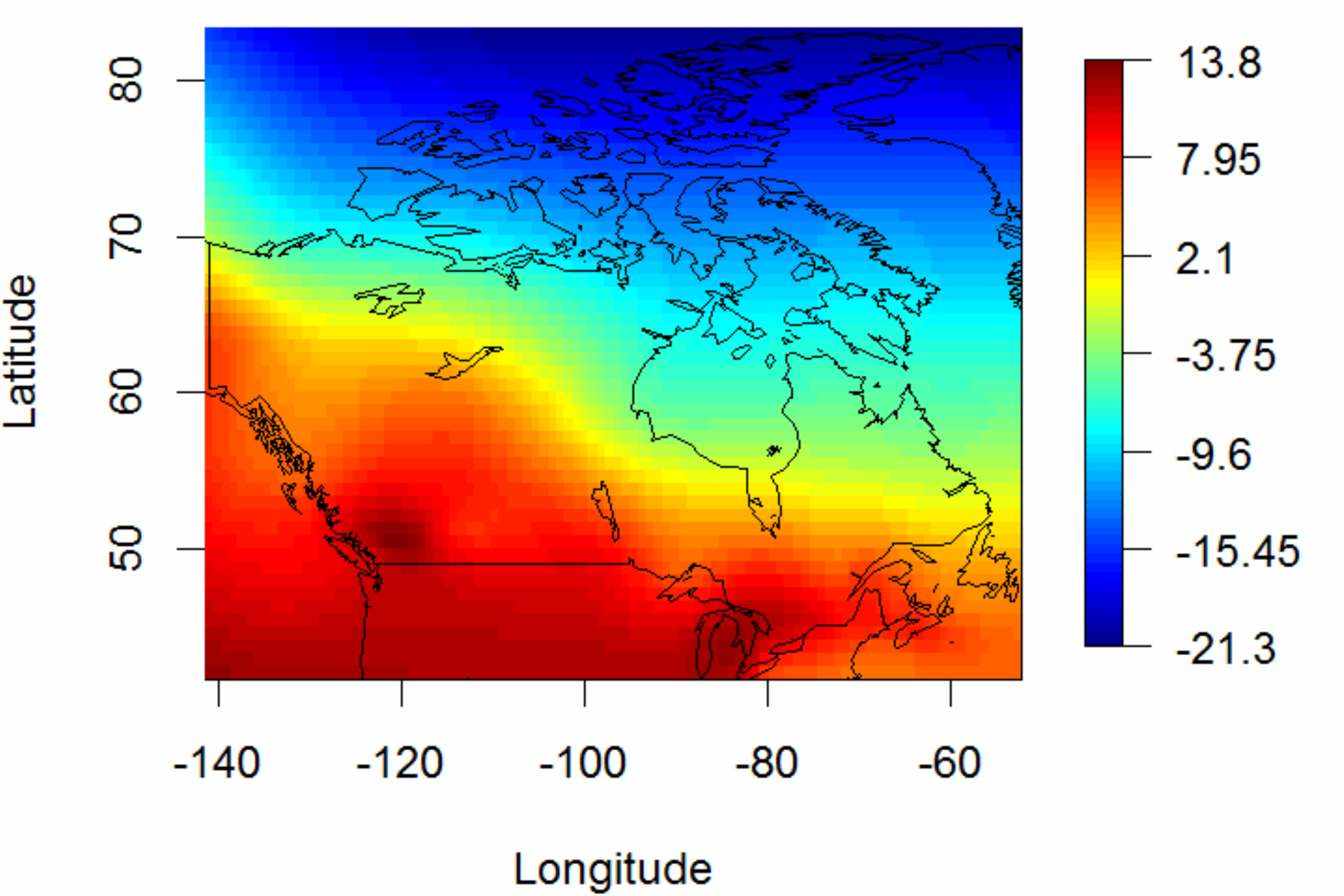}\tabularnewline
(b1) & (b2)\tabularnewline
\includegraphics[scale=0.32]{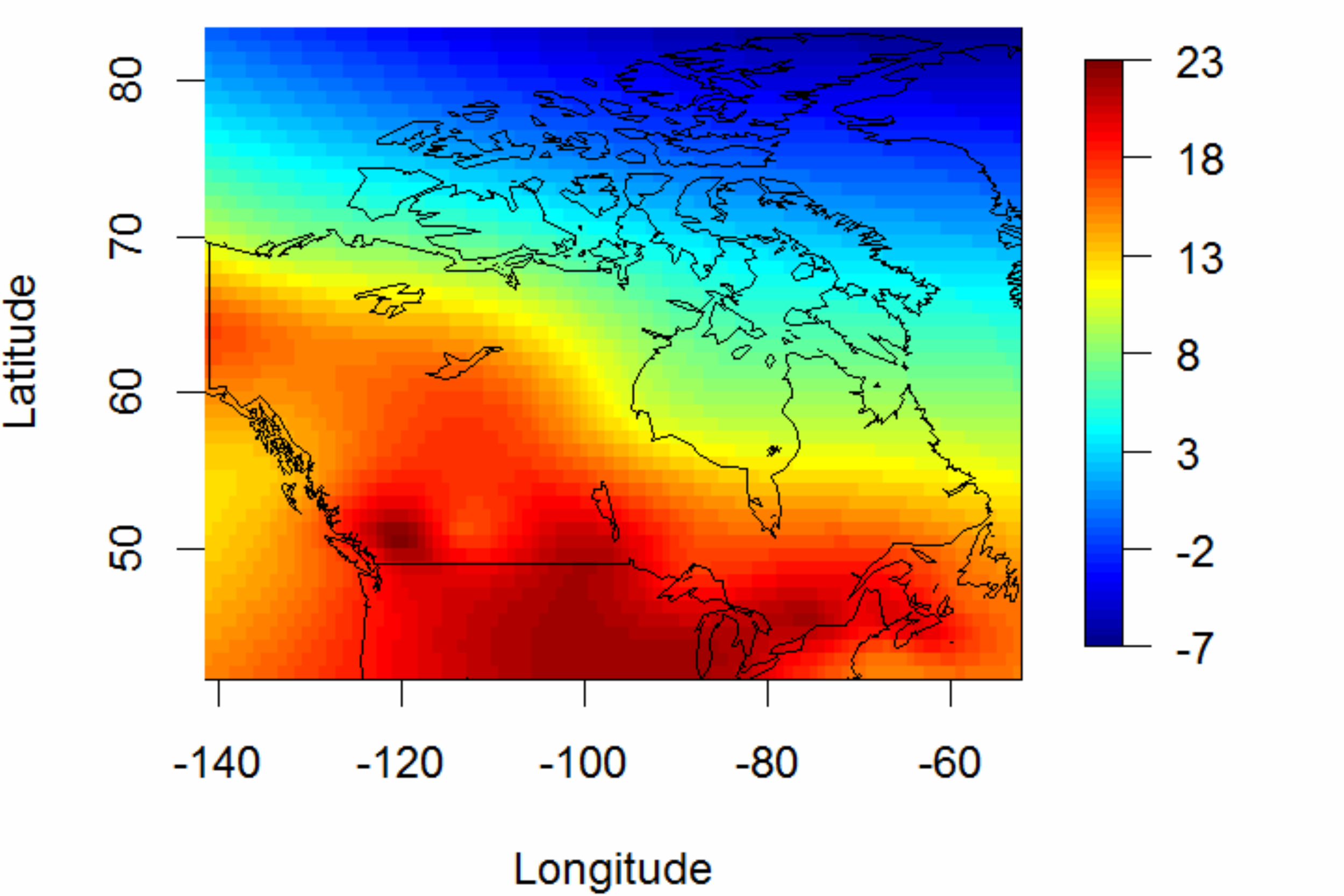}  & \includegraphics[scale=0.32]{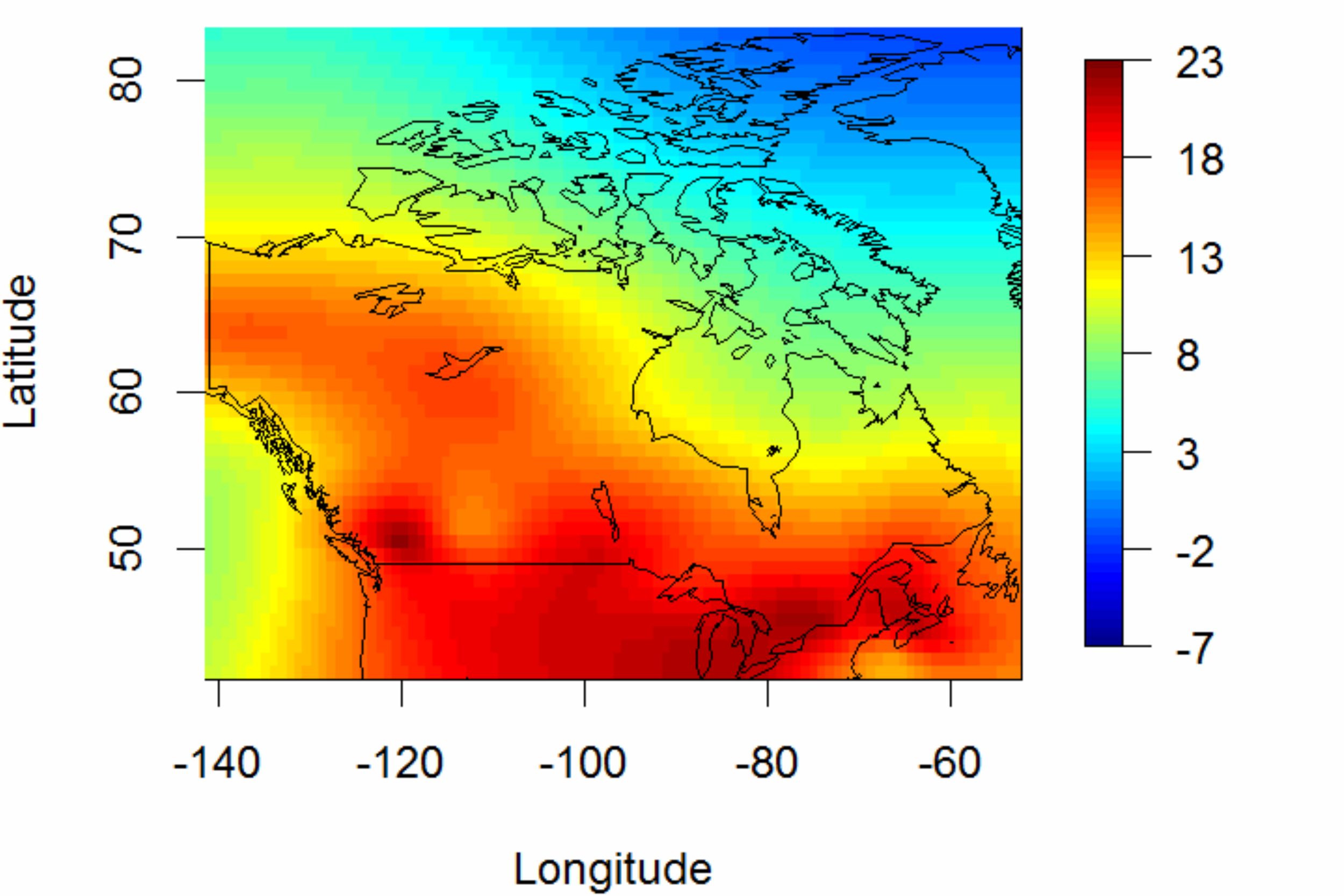}\tabularnewline
(c1) & (c2)\tabularnewline
\end{tabular}
\caption{(a1) $\mu(\bm{s},50)$; (a2) $\hat{y}(\bm{s},50)$; (b1) $\mu(\bm{s},125)$;
	(b2) $\hat{y}(\bm{s},125)$; (c1) $\mu(\bm{s},200)$; (c2) $\hat{y}(\bm{s},200)$.}
\label{fig:prediction map} 
\end{figure}

The performance of the two covariance function estimates is evaluated in terms of the Frobenius loss,
$\mathrm{Loss}_{F}=\|\hat{\bm{\Sigma}}-\bm{S}_{\mathrm{test}}\|$ and the Kullbeck-Leibler loss,
$\mathrm{Loss}_{KL}=\frac{1}{2}\big\{\mathrm{tr}(\hat{\bm{\Sigma}}^{-1}\bm{S}_{\mathrm{test}})
+\log|\hat{\bm{\Sigma}}|-\log|\bm{S}_{\mathrm{test}}|-35\big\}$, where
$\bm{\hat{\Sigma}}$ is a generic estimate of $\bm{\Sigma}$ and $\bm{S}_{\mathrm{test}}$
is the sample covariance matrix based on the testing data. The validation procedure was repeated 100 times.
The average $\mathrm{Loss}_F$ and $\mathrm{Loss}_{KL}$ based on our method are $10.0$ and $4.7$ respectively, which are
much smaller than $177.1$ and $25.7$ based on the exponential covariance model, which is not surprising,
because the data are highly nonstationary in space.
The mean surfaces $\mu(\cdot,t)$ and the final predicted surfaces $\hat{y}(\cdot,t)$ of $y(\cdot,t)$
for $t=50,125,200$ are shown in Figure~\ref{fig:prediction map}. 

\section*{Appendix}

\paragraph{Proof of Theorem 1}

(i) We first show that $\displaystyle\sum_{k=1}^{n}a_{k}f_{k}(\cdot)\in\mathcal{F}$, for any given $a_1,\dots,a_n\in\mathbb{R}$.
Direct calculation gives
\[
\sum_{k=1}^{n}a_{k}f_{k}(\cdot)=\bm{\alpha}'\bm{\phi}(\bm{s})+\bm{\beta}'(1,x_{1},\dots,x_{d})',
\]
where
\begin{align}
	\bm{\alpha}
=&~ \bm{V}_{n-d-1}\mathrm{diag}(\lambda_{1}^{-1},\dots,\lambda_{n-d-1}^{-1})(a_{d+2},\dots,a_{n})',
	\label{eq:alpha for g}\\
	\bm{\beta}
=&~ (a_{1},\ldots,a_{d+1})'-(\bm{X}^{\prime}\bm{X})^{-1}\bm{X}'\bm{\Phi}\bm{V}_{n-d-1}
	\mathrm{diag}(\lambda_{1}^{-1},\dots,\lambda_{n-d-1}^{-1})(a_{d+2},\dots,a_{n})',\nonumber 
\end{align}
and $\bm{V}_{n-d-1}$ is the submatrix of $\bm{V}$ in (\ref{eq:basis detail}) consisting
of its first $n-d-1$ columns. By the definition of $\bm{V}$,
$\bm{Q}\bm{\Phi}\bm{Q}=\bm{V}_{n-d-1}\mathrm{diag}(\lambda_{1},\ldots,\lambda_{n-d-1})\bm{V}_{n-d-1}'$, and hence
\begin{equation}
\bm{V}_{n-d-1}=\bm{Q}\bm{\Phi}\bm{Q}\bm{V}_{n-d-1}\mathrm{diag}(\lambda_{1}^{-1},\ldots,\lambda_{n-d-1}^{-1}).
\label{eq:partial V with positive eigenvalues}
\end{equation}
		
\noindent This together with \eqref{eq:alpha for g} and $\bm{X}'\bm{Q}=\bm{0}$ implies that $\bm{X}'\bm{\alpha}=\bm{0}$.
Thus $\displaystyle\sum_{k=1}^{n}a_{k}f_{k}(\cdot)\in\mathcal{F}$ is proved.

We remain to show that $\mathcal{F}\subset\Big\{\displaystyle\sum_{k=1}^{n}a_k f_k(\cdot):a_k\in\mathbb{R}\Big\}$.
We first show that
\begin{equation}
\bm{V}_{n-d-1}\bm{V}_{n-d-1}'=\bm{Q}.
\label{eq:V is half of Q}
\end{equation}

\noindent From \eqref{eq:partial V with positive eigenvalues} and $\bm{X}'\bm{Q}=\bm{0}$,
we have $\bm{X}\bm{V}_{n-d-1}\bm{V}_{n-d-1}'=\bm{0}$. This and the fact that
$\bm{V}_{n-d-1}\bm{V}_{n-d-1}'$ is idempotent of rank $n-d-1$ imply that $\bm{V}_{n-d-1}\bm{V}_{n-d-1}'$ is
the projection matrix for the space orthogonal to the column space of $\bm{X}$. That is, $\bm{V}_{n-d-1}\bm{V}_{n-d-1}'=\bm{Q}$.

Given any $f(\bm{s})=\bm{\alpha}'\bm{\phi}(\bm{s})+\beta_{0}+\displaystyle\sum_{j=1}^{d}\beta_{j}x_{j}\in\mathcal{F}$,
since $\bm{X}'\bm{\alpha}=\bm{0}$, we can write
\begin{align*}
	f(\bm{s})
=&~ \bm{\phi}(\bm{s})'(\bm{\alpha}-\bm{X}(\bm{X}'\bm{X})^{-1}\bm{X}'\bm{\alpha})+(1,x_1,\dots,x_d)'\bm{\beta}\\
=&~ (\bm{\phi}(\bm{s})',1,x_1,\dots,x_d)
	\begin{bmatrix}
	\bm{V}_{n-d-1}\bm{V}_{n-d-1}' & \bm{0}\\
	\bm{0} & \bm{I}_{d+1}
	\end{bmatrix}
	\begin{bmatrix}
	\bm{\alpha}\\
	\bm{\beta}
	\end{bmatrix}\\
=&~ (\bm{\phi}(\bm{s})^{\prime},1,x_{1},\ldots,x_{d})
	\begin{bmatrix}
	\bm{0} & \bm{V}_{n-d-1}\mathrm{diag}(\lambda_{1}^{-1},\ldots,\lambda_{n-d-1}^{-1})\\
	\bm{I}_{d+1} & -(\bm{X}^{\prime}\bm{X})^{-1}\bm{X}^{\prime}\bm{\Phi}\bm{V}_{n-d-1}\mathrm{diag}(\lambda_{1}^{-1},\ldots,\lambda_{n-d-1}^{-1})
	\end{bmatrix}\\
&~ \times
	\begin{bmatrix}
	(\bm{X}^{\prime}\bm{X})^{-1}\bm{X}^{\prime}\bm{\Phi}\bm{V}_{n-d-1}\bm{V}_{n-d-1}^{\prime} & \bm{I}_{d+1}\\
	\mathrm{diag}(\lambda_{1},\ldots,\lambda_{n-d-1})\bm{V}_{n-d-1}^{\prime} & \bm{0}
	\end{bmatrix}
	\begin{bmatrix}
	\bm{\alpha}\\
	\bm{\beta}
	\end{bmatrix}\\
=&~ (f_{1}(\bm{s}),\ldots,f_{n}(\bm{s}))
	\begin{bmatrix}
	(\bm{X}^{\prime}\bm{X})^{-1}\bm{X}^{\prime}\bm{\Phi}\bm{V}_{n-d-1}\bm{V}_{n-d-1}^{\prime}\bm{\alpha}+\bm{\beta}\\
	\mathrm{diag}(\lambda_{1},\ldots,\lambda_{n-d-1})\bm{V}_{n-d-1}^{\prime}\bm{\alpha}
	\end{bmatrix},
\end{align*}

\noindent where the second equality follows from \eqref{eq:V is half of Q}. Thus $f(\cdot)\in
\Big\{\displaystyle\sum_{k=1}^{n}a_k f_k(\cdot):a_k\in\mathbb{R}\Big\}$. This completes the proof of (i).

(ii) Clearly, $J(f_1)=\cdots=J(f_{d+1})=0$. It suffices to show that $J(f)=\bm{\alpha}'\bm{\Phi}\bm{\alpha}>0$ for any $f(\bm{s})=\bm{\alpha}'\bm{\phi}(\bm{s})+\beta_0
+\displaystyle\sum_{j=1}^d\beta_j x_j\in\mathcal{F}$ with $\bm{\alpha}\neq\bm{0}$. Since $\mathrm{rank}(\bm{X})=d+1$,
$\bm{X}'\bm{\alpha}=\bm{0}$ and $\bm{\alpha}\neq\bm{0}$, it follows that
$\bm{\alpha}'\bm{\Phi}\bm{\alpha}>0$ (see Section 4 of \citet{Micchelli1986}). This completes the proof of (ii).

(iii) We shall only prove the result for $k=d+2$. Given any $g(\cdot)\in\mathcal{F}$, let $\bm{g}=(g(\bm{s}_1),\dots,g(\bm{s}_{n}))'=
\bm{\Phi}\bm{\alpha}_g+\bm{X}\bm{\beta}_g$. Then $g(\cdot)\in\mathcal{F}_{d+2}$ if and only if
\begin{align}
	(\bm{\alpha}_g,\bm{\beta}_g)
\in&~ \big\{(\bm{\alpha},\bm{\beta}):\bm{X}'\bm{\alpha}=0,\,\bm{X}'\bm{g}=\bm{0},\mbox{ and }\|\bm{g}\|_2=1\big\}\nonumber\\
=&~ \big\{(\bm{\alpha},\bm{\beta}):\bm{X}'\bm{\alpha}=0,\,\bm{g}=\bm{Q}\bm{g}=\bm{Q}\bm{\Phi}\bm{\alpha},\mbox{ and }\|\bm{g}\|_2=1\big\}\nonumber\\
=&~ \big\{(\bm{\alpha},\bm{\beta}):\bm{\alpha}=\bm{Q}\bm{\alpha},\,\bm{\beta}=-(\bm{X}'\bm{X})^{-1}\bm{X}'\bm{\Phi}\bm{\alpha},
	\mbox{ and }\|\bm{Q}\bm{\Phi}\bm{Q}\bm{\alpha}\|_2=1\big\}.
\label{eq:beta hat}
\end{align}
Therefore, from \eqref{eq:quadratic loss} and \eqref{eq:beta hat},
\begin{align}
	\min_{g(\cdot)\in\mathcal{F}_{d+2}}J(g)
=&~ \min\{\bm{\alpha}'\bm{\Phi}\bm{\alpha}:\bm{\alpha}\in\mathbb{R}^n,\,\bm{\alpha}=\bm{Q}\bm{\alpha},\,\|\bm{Q}\bm{\Phi}\bm{Q}\bm{\alpha}\|_2=1\}\nonumber\\
=&~ \min\{\bm{\alpha}'\bm{Q}\bm{\Phi}\bm{Q}\bm{\alpha}:\bm{\alpha}\in\mathbb{R}^n,\,\|\bm{Q}\bm{\Phi}\bm{Q}\bm{\alpha}\|_2=1\}\nonumber\\
=&~ \min\{\bm{\alpha}'\bm{V}\bm{\Lambda}\bm{V}'\bm{\alpha}:\bm{\alpha}\in\mathbb{R}^n,\,
	\|\bm{\Lambda}\bm{V}'\bm{\alpha}\|_2=1\}\nonumber\\
=&~ \min\{\bm{a}'\bm{\Lambda}\bm{a}:\bm{a}\in\mathbb{R}^n,\,\|\bm{\Lambda}\bm{a}\|_2=1\}=\lambda_1^{-1},
\label{eq:f1}
\end{align}

\noindent where $\bm{\Lambda}=\mathrm{diag}(\lambda_1,\dots,\lambda_n)$. It follows from \eqref{eq:beta hat} and \eqref{eq:f1} that
\[
\big(\lambda_1^{-1}\bm{v}_1,\,-\lambda_1^{-1}(\bm{X}'\bm{X})^{-1}\bm{X}'\bm{\Phi}\bm{v}_1\big)=\mathop{\arg\min}_{(\bm{\alpha},\bm{\beta})}
\big\{J(g):g(\bm{x})=\bm{\phi}(\bm{x})'\bm{\alpha}+(1,x_2,\dots,x_d)'\bm{\beta}\in\mathcal{F}_{d+2}\big\}.
\]
This proves (iii) and the proof of Theorem 2 is complete.

\paragraph{Proof of Theorem 2}

Let $\bm{H}=\bm{F}_K(\bm{F}'_K\bm{F}_K)^{-1}\bm{F}'_K$, $\bm{L}_K=\bm{F}_K(\bm{F}'_K\bm{F}_K)^{-1/2}$, and
$\bm{R}=\bm{L}_K\bm{P}_K$. It follows from the definition of $\bm{P}_K\mathrm{diag}(d_{K,1},\dots,d_{K,K})\bm{P}'_K$
and simple algebra that $\bm{H}\bm{S}\bm{H}=\bm{R}\,\mathrm{diag}(d_{K,1},\dots,d_{K,K})\bm{R}'$.
Since $\mathrm{rank}(\bm{F}_K\bm{M}\bm{F}'_K)\leq K$, the eigen-decomposition of $\bm{F}_K\bm{M}\bm{F}'_K$
can be written as $\tilde{\bm{R}}\,\mathrm{diag}(\tilde{d}_1,\dots,\tilde{d}_K)\tilde{\bm{R}}'$, where $\tilde{\bm{R}}$ is an $n\times K$ matrix
with orthonormal columns. Using $\bm{H}\bm{F}_K=\bm{F}_K$, we have
\begin{align*}
	\{\bm{F}_K\bm{M}
& \bm{F}'_K+(\sigma_\xi^2+\sigma_\epsilon^2)\bm{I}_{n}\}^{-1}\\
=&~ \big\{\bm{H}\tilde{\bm{R}}\,\mathrm{diag}\big(\tilde{d}_1,\dots,\tilde{d}_K\big)\tilde{\bm{R}}'\bm{H}+
	(\sigma_\xi^2+\sigma_\epsilon^2)\bm{I}_{n}\big\}^{-1}\\
=&~ \frac{1}{\sigma_\xi^2+\sigma_\epsilon^2}\bm{I}_n-\frac{1}{\sigma_\xi^2+\sigma_\epsilon^2}\bm{H}\tilde{\bm{R}}
	\,\mathrm{diag}\bigg(\frac{\tilde{d}_1}{\tilde{d}_1+\sigma_\xi^2+\sigma_\epsilon^2},\dots,
	\frac{\tilde{d}_K}{\tilde{d}_K+\sigma_\xi^2+\sigma_\epsilon^2}\bigg)\tilde{\bm{R}}'\bm{H}.
\end{align*}

\noindent Then twice the negative log-likelihood function of $\bm{z}_1,\dots,\bm{z}_T$ is
\begin{align}
\ell(\bm{M},
& \sigma_\xi^2)=nT\log2\pi+\log\big|\bm{F}_K\bm{M}\bm{F}'_K+(\sigma_\xi^2+\sigma_\epsilon^2)\bm{I}_n\big|+\mathrm{tr}\big\{\bm{S}
	(\bm{F}_K\bm{M}\bm{F}'_K+(\sigma_{\xi}^{2}+\sigma_\epsilon^2)\bm{I}_{n}\big)^{-1}\big\}\notag\\
=&~ nT\log2\pi+\log\big|\tilde{\bm{R}}\,\mathrm{diag}\big(\tilde{d}_1,\dots,\tilde{d}_K\big)\tilde{\bm{R}}'+(\sigma_\xi^2+\sigma_\epsilon^2)\bm{I}_{n}\big|\notag\\
&~ +\mathrm{tr}\bigg(\frac{1}{\sigma_\xi^2+\sigma_\epsilon^2}\bm{S}-\frac{1}{\sigma_\xi^2+\sigma_\epsilon^2}\bm{S}\bm{H}\tilde{\bm{R}}
	\,\mathrm{diag}\bigg(\frac{\tilde{d}_1}{\tilde{d}_1+\sigma_\xi^2+\sigma_\epsilon^2},\dots,
	\frac{\tilde{d}_K}{\tilde{d}_K+\sigma_\xi^2+\sigma_\epsilon^2}\bigg)\tilde{\bm{R}}'\bm{H}\bigg)\notag\\
=&~ nT\log2\pi+\bigg\{\sum_{k=1}^K\log\big(\tilde{d}_k+\sigma_\xi^2+\sigma_\epsilon^2\big)\bigg\}+(n-K)\log(\sigma_\xi^2+\sigma_\epsilon^2)
	+\frac{1}{\sigma_{\xi}^{2}+\sigma_{\epsilon}^{2}}\mathrm{tr}(\bm{S})\notag\\
&~ -\frac{1}{\sigma_\xi^2+\sigma_\epsilon^2}\mathrm{tr}\bigg(\bm{R}\,\mathrm{diag}(d_{K,1},\dots,d_{K,K})\bm{R}'
	\tilde{\bm{R}}\,\mathrm{diag}\bigg(\frac{\tilde{d}_1}{\tilde{d}_1+\sigma_\xi^2+\sigma_\epsilon^2},\dots,
	\frac{\tilde{d}_K}{\tilde{d}_K+\sigma_\xi^2+\sigma_\epsilon^2}\bigg)\tilde{\bm{R}}'\bigg)\notag\\
\geq&~ nT\log2\pi+\bigg\{\sum_{k=1}^K\log\big(\tilde{d}_k+\sigma_\xi^2+\sigma_\epsilon^2\big)\bigg\}+(n-K)\log(\sigma_\xi^2+\sigma_\epsilon^2)
	+\frac{1}{\sigma_{\xi}^{2}+\sigma_{\epsilon}^{2}}\mathrm{tr}(\bm{S}) \nonumber\\
&~ -\frac{1}{\sigma_\xi^2+\sigma_\epsilon^2}\sum_{k=1}^K\frac{d_{K,k}\tilde{d}_k}{\tilde{d}_k+\sigma_\xi^2+\sigma_\epsilon^2}\:,
\label{eq:loglik}
\end{align}
where the last inequality follows from von Neumann's trace inequality (\citealp{neumann1937some})
and the equality holds if and only if $\bm{\tilde{R}}=\bm{R}$. So given $\sigma_\xi^2$,
$\ell(\bm{M},\sigma_\xi^2)$ is minimized at $\hat{\bm{M}}_K(\sigma_\xi^2)$ such that
$\bm{F}_K\hat{\bm{M}}_K\bm{F}'_K=\bm{R}\,\mathrm{diag}
\big(\hat{d}_{K,1}(\sigma_\xi^2),\dots,\hat{d}_{K,K}(\sigma_\xi^2)\big)\bm{R}'$, where
$\hat{d}_{K,k}(\sigma_\xi^2)=\max(d_{K,k}-\sigma_\xi^2-\sigma_\epsilon^2,\,0)$; $k=1,\dots,K$. It follow that
\begin{align*}
	\hat{\bm{M}}_K
=&~ (\bm{F}'_K\bm{F}_K)^{-1}\bm{F}'_K\bm{F}_K\hat{\bm{M}}_K\bm{F}'_K\bm{F}_K(\bm{F}'_K\bm{F}_K)^{-1}\\
=&~ (\bm{F}'_K\bm{F}_K)^{-1}\bm{F}'_K\bm{R}\,\mathrm{diag}\big(\hat{d}_{K,1},\dots,\hat{d}_{K,K}\big)\bm{R}'\bm{F}_K(\bm{F}'_K\bm{F}_K)^{-1}\\
=&~ (\bm{F}'_K\bm{F}_K)^{-1/2}\bm{P}_K\,\mathrm{diag}\big(\hat{d}_{K,1},\dots,\hat{d}_{K,K}\big)\bm{P}_K(\bm{F}'_K\bm{F}_K)^{-1/2}.
\end{align*}
Finally, replacing $\tilde{d}_k$ in the righthand side of \eqref{eq:loglik} by $\hat{d}_k$,
for $k=1,\dots,K$, we obtain the desired result for $\hat{\sigma}_{\xi,K}^2$.
This completes the proof. 

\bibliography{reference}

\begin{thebibliography}{}

\bibitem[\protect\citeauthoryear{Akaike}{Akaike}{1973}]{akaike1973}
Akaike, H. (1973).
\newblock Information theory and an extension of the maximum likelihood
  principle.
\newblock In B.~Petro and F.~Cs{\'a}ki (Eds.), {\em Proceedings of the Second
  International Symposium on Information Theory}, pp.\  267--281. Budapest:
  Akad{\'e}miai Kiad{\'o}.

\bibitem[\protect\citeauthoryear{Akaike}{Akaike}{1974}]{akaike1974}
Akaike, H. (1974).
\newblock A new look at the statistical model identification.
\newblock {\em Automatic Control, IEEE Transactions on\/}~{\em 19}, 716--723.

\bibitem[\protect\citeauthoryear{Barry, Jay, and Hoef}{Barry
  et~al.}{1996}]{barry1996blackbox}
Barry, R.~P., M.~Jay, and V.~Hoef (1996).
\newblock Blackbox kriging: spatial prediction without specifying variogram
  models.
\newblock {\em Journal of Agricultural, Biological, and Environmental
  Statistics\/}~{\em 5}, 297--322.

\bibitem[\protect\citeauthoryear{Buja, Hastie, and Tibshirani}{Buja
  et~al.}{1989}]{buja1989linear}
Buja, A., T.~Hastie, and R.~Tibshirani (1989).
\newblock Linear smoothers and additive models.
\newblock {\em The Annals of Statistics\/}~{\em 17}, 453--510.

\bibitem[\protect\citeauthoryear{Cressie and Johannesson}{Cressie and
  Johannesson}{2008}]{cressie2008fixed[clm]}
Cressie, N. and G.~Johannesson (2008).
\newblock Fixed rank kriging for very large spatial data sets.
\newblock {\em Journal of the Royal Statistical Society: Series B\/}~{\em 70},
  209--226.

\bibitem[\protect\citeauthoryear{Demmler and Reinsch}{Demmler and
  Reinsch}{1975}]{demmler1975oscillation[TPS]}
Demmler, A. and C.~Reinsch (1975).
\newblock Oscillation matrices with spline smoothing.
\newblock {\em Numerische Mathematik\/}~{\em 24}, 375--382.

\bibitem[\protect\citeauthoryear{Golub, Heath, and Wahba}{Golub
  et~al.}{1979}]{golub1979generalized}
Golub, G.~H., M.~Heath, and G.~Wahba (1979).
\newblock Generalized cross-validation as a method for choosing a good ridge
  parameter.
\newblock {\em Technometrics\/}~{\em 21}, 215--223.

\bibitem[\protect\citeauthoryear{Golub and van~der Vorst}{Golub and van~der
  Vorst}{2000}]{golub2000eigenvalue}
Golub, G.~H. and H.~A. van~der Vorst (2000).
\newblock Eigenvalue computation in the 20th century.
\newblock {\em Journal of Computational and Applied Mathematics\/}~{\em 123},
  35--65.

\bibitem[\protect\citeauthoryear{Green and Silverman}{Green and
  Silverman}{1993}]{green1993nonparametric}
Green, P.~J. and B.~W. Silverman (1993).
\newblock {\em Nonparametric regression and generalized linear models: a
  roughness penalty approach}.
\newblock CRC Press.

\bibitem[\protect\citeauthoryear{Hastie and Tibshirani}{Hastie and
  Tibshirani}{1990}]{hastie1990generalized}
Hastie, T.~J. and R.~J. Tibshirani (1990).
\newblock {\em Generalized additive models}.
\newblock CRC Press.

\bibitem[\protect\citeauthoryear{Katzfuss and Cressie}{Katzfuss and
  Cressie}{2009}]{katzfuss2009EM[clm]}
Katzfuss, M. and N.~Cressie (2009).
\newblock Maximum likelihood estimation of covariance parameters in the
  spatial- random-effects model.
\newblock In {\em 2009 Proceedings of the Joint Statistical Meetings}, pp.\
  3378--3390. Alexandria, VA: American Statistical Association.

\bibitem[\protect\citeauthoryear{Lemos and Sans{\'o}}{Lemos and
  Sans{\'o}}{2012}]{lemos2012conditionally[clm-review]}
Lemos, R.~T. and B.~Sans{\'o} (2012).
\newblock Conditionally linear models for non-homogeneous spatial random
  fields.
\newblock {\em Statistical Methodology\/}~{\em 9}, 275--284.

\bibitem[\protect\citeauthoryear{Micchelli}{Micchelli}{1986}]{Micchelli1986}
Micchelli, C.~A. (1986).
\newblock Interpolation of scattered data: Distance matrices and conditionally
  positive definite functions.
\newblock {\em Constructive Approximation\/}~{\em 2}, 11--22.

\bibitem[\protect\citeauthoryear{Nychka, Bandyopadhyay, Hammerling, Lindgren,
  and Sain}{Nychka et~al.}{2015}]{nychka2014multi}
Nychka, D., S.~Bandyopadhyay, D.~Hammerling, F.~Lindgren, and S.~Sain (2015).
\newblock A multi-resolution gaussian process model for the analysis of large
  spatial data sets.
\newblock {\em Journal of Computational and Graphical Statistics\/}~{\em to
  appear}.

\bibitem[\protect\citeauthoryear{Nychka, Wikle, and Royle}{Nychka
  et~al.}{2002}]{nychka2002}
Nychka, D., C.~Wikle, and J.~A. Royle (2002).
\newblock Multiresolution models for nonstationary spatial covariance
  functions.
\newblock {\em Statistical Modelling\/}~{\em 2\/}(4), 315--331.

\bibitem[\protect\citeauthoryear{Ordonez, Mohanam, and Garcia-Alvarado}{Ordonez
  et~al.}{2014}]{ordonez2014pca}
Ordonez, C., N.~Mohanam, and C.~Garcia-Alvarado (2014).
\newblock Pca for large data sets with parallel data summarization.
\newblock {\em Distributed and Parallel Databases\/}~{\em 32}, 377--403.

\bibitem[\protect\citeauthoryear{Ramsay and Dalzell}{Ramsay and
  Dalzell}{1991}]{ramsay1991some}
Ramsay, J.~O. and C.~Dalzell (1991).
\newblock Some tools for functional data analysis.
\newblock {\em Journal of the Royal Statistical Society. Series B\/}~{\em 53},
  539--572.

\bibitem[\protect\citeauthoryear{Sampson and Guttorp}{Sampson and
  Guttorp}{1992}]{sampson1992nonparametric[deformation]}
Sampson, P.~D. and P.~Guttorp (1992).
\newblock Nonparametric estimation of nonstationary spatial covariance
  structure.
\newblock {\em Journal of the American Statistical Association\/}~{\em 87},
  108--119.

\bibitem[\protect\citeauthoryear{Shi and Cressie}{Shi and
  Cressie}{2007}]{shi2007global}
Shi, T. and N.~Cressie (2007).
\newblock Global statistical analysis of misr aerosol data: a massive data
  product from nasa's terra satellite.
\newblock {\em Environmetrics\/}~{\em 18\/}(7), 665--680.

\bibitem[\protect\citeauthoryear{Silverman and Ramsay}{Silverman and
  Ramsay}{2005}]{silverman2005functional}
Silverman, B. and J.~Ramsay (2005).
\newblock {\em Functional Data Analysis}.
\newblock Springer.

\bibitem[\protect\citeauthoryear{Silverman}{Silverman}{1995}]{silverman1995incorporating}
Silverman, B.~W. (1995).
\newblock Incorporating parametric effects into functional principal components
  analysis.
\newblock {\em Journal of the Royal Statistical Society. Series B\/}~{\em 57},
  673--689.

\bibitem[\protect\citeauthoryear{von Neumann}{von
  Neumann}{1937}]{neumann1937some}
von Neumann, J. (1937).
\newblock Some matrix inequalities and metrization of matrix space.
\newblock {\em Tomsk Universitet Review\/}~{\em 1}, 286--300.

\bibitem[\protect\citeauthoryear{Wahba and Wendelberger}{Wahba and
  Wendelberger}{1980}]{wahba1980some[TPS]}
Wahba, G. and J.~Wendelberger (1980).
\newblock Some new mathematical methods for variational objective analysis
  using splines and cross validation.
\newblock {\em Monthly weather review\/}~{\em 108}, 1122--1143.

\bibitem[\protect\citeauthoryear{Wikle}{Wikle}{2010}]{wikle2010[clm]}
Wikle, C. (2010).
\newblock Low-rank representations for spatial processes.
\newblock In M.~F. A.~E.~Gelfand, P. J.~Diggle and P.~Guttorp (Eds.), {\em
  Handbook of Spatial Statistics}, pp.\  107--118. CRC Press, Boca Raton,
  Florida, USA.

\end{thebibliography}
\end{document}